\documentclass[%
 reprint,
 amsmath,amssymb,
 aps,
]{revtex4-1}

\usepackage{graphicx}
\usepackage{dcolumn}
\usepackage{bm}
\usepackage{color}

\newcommand {\e}[1]{\mathrm{~#1}}       
\newcommand {\vek}[1]{\mathbf{#1}}

\newcommand{\figref}[1]{Fig~\ref{#1}}

\renewcommand{\eqref}[1]{Eq.~(\ref{#1})}

\newcommand{\be}{\begin{equation}}
\newcommand{\ee}{\end{equation}}

\newcommand{\la}{\lambda}

\begin{document}

\preprint{APS/123-QED}

\title{Statistical mechanics for  metabolic networks during steady-state growth.}

\author{Daniele De Martino}
\author{Anna MC Andersson}%
\author{Tobias Bergmiller}
\author{C\u{a}lin C Guet}
\author{Ga\v{s}per Tka\v{c}ik}

\affiliation{%
 Institute of Science and Technology Austria,\\ Am Campus 1, A-3400 Klosterneuburg, Austria
}%





\begin{abstract}
Which properties of metabolic networks can be derived solely from stoichiometric information about the network's constituent reactions? 
Predictive results have been obtained by Flux Balance Analysis (FBA), by postulating that cells set metabolic fluxes within the allowed stoichiometry so as to maximize their growth. Here, we generalize this framework   to single cell level using maximum entropy models from statistical physics. We define and compute, for the core metabolism of \emph{Escherichia coli}, a joint distribution over all  fluxes that yields the experimentally observed growth rate. This solution, containing FBA as a limiting case, provides a better match to the measured fluxes in the wild type and several mutants. We find that \emph{E. coli} metabolism is close to, but not at, the optimality assumed by FBA.
 Moreover, our model makes a wide range of predictions: (i) on flux variability, its regulation, and flux correlations across individual cells; (ii) on the relative importance of stoichiometric constraints vs. growth rate optimization; (iii) on  quantitative scaling relations for singe-cell growth rate distributions. We validate these scaling predictions using data from individual bacterial cells grown in a microfluidic device at different sub-inhibitory antibiotic concentrations. Under mild dynamical assumptions, fluctuation-response relations  further predict the autocorrelation timescale  in growth data and growth rate adaptation times following an environmental perturbation.
\end{abstract}

\pacs{Valid PACS appear here}
\maketitle

{\bf Significance: }{\em While metabolic reactions have been studied in detail in model organisms and can today in principle be inferred  automatically from genome-scale data, our systems-level understanding of the complete metabolic networks rests primarily on flux balance analysis (FBA). In FBA, known reactions impose physical constraints within which free parameters are chosen to maximize the growth rate. Using ideas from statistical physics, we extend this approach to capture  cell-to-cell variability. Our model is easy to infer, provides superior fit to measurements in \emph{Escherichia coli}, and makes several categorically new predictions connecting metabolism to single cell physiology. We find that \emph{E. coli} grows close to, but not at, optimality, and further predict and experimentally test the scaling of growth rate fluctuations in clonal bacterial populations.} 
\section{Introduction}
After the significant developments of molecular biology and biochemistry of the last century, many aspects of cellular physiology could be understood as a result of interactions between  identified molecular components. Perhaps the best-characterized example is intermediate metabolism, the set of reactions that enable cell growth by converting organic compounds and transducing free energy. Today, one can infer nearly complete metabolic networks from genome-scale data, but the dynamics and parameter dependence of such networks remain difficult to analyze. Alternatively, one can assume that known reactions only provide physico-chemical constraints within which some adaptive dynamics has maximized the growth rate, e.g., by adjusting enzyme levels  and controlling reaction rates \cite{kacser1995control}. An influential implementation of this idea for batch cultures under steady state conditions has been the flux balance analysis (FBA)~\cite{Orth:2010if}, which has been tested experimentally~\cite{ibarra2002escherichia, edwards2001silico}, also in mutant strains, strains used for industrial production~\cite{majewski1990simple, Varma:1994gd, Edwards:2000zt}, as well as  phenotypes implicated in disease (e.g., Warburg effect~\cite{vazquez2010catabolic}). Using maximum entropy ideas from statistical physics, we extend the application of FBA from batch to single-cell level and show that our extension makes a wide range of predictions, some of which we test experimentally.

Recent measurements at the single-cell level demonstrated the existence of substantial cell-to-cell growth rate fluctuations even in well-controlled steady-state conditions~\cite{wang2010robust}. These fluctuations exhibit universal scaling properties~\cite{iyer2014universality, iyer2014scaling, kennard2016individuality}, relate to cell size control mechanisms~\cite{taheri2015cell}, act as a global collective mode for heterogeneity  in gene expression~\cite{kiviet2014stochasticity, shahrezaei2015connecting, keren2015noise}, and are ultimately believed to affect fitness~\cite{cerulus2016noise}. To link these observations to metabolism, however, we need to set up a mathematical description not only of the optimal metabolic fluxes and maximal growth rate in batch culture (as in FBA, which permits no heterogeneity across cells), but for the complete joint distribution over metabolic fluxes. Metabolic phenotypes of individual cells growing in steady-state conditions can then be understood as samples from this joint distribution, which would automatically contain information about flux correlations, and, in particular, could directly predict cell-to-cell growth rate fluctuations.

The simplest construction of a joint distribution over metabolic fluxes can be derived in the maximum entropy framework~\cite{jaynes1957information}. The key intuition is to look for the most unbiased (or random) distribution over fluxes through individual metabolic reactions that is consistent with the given stoichiometric constraints, while matching the experimentally measured average growth rate. The maximum entropy model that we specify below will turn out to be a one-parameter family of distributions, where the single parameter can be fit to match experimental data; all subsequent predictions follow directly, without further fitting. 
A similar approach has recently been used in diverse biological settings, ranging from neural networks~\cite{schneidman2006weak,tkacikpcb}, genetic regulatory networks~\cite{lezon2006using}, antibody diversity~\cite{mora2010maximum}, and collective motion of starling flocks~\cite{bialek2012statistical}.

In addition to accounting for fluctuations, the maximum entropy construction provides a principled interpolation between  two extremal regimes of metabolic network function. In the  ``uniform'' (no-optimization) limit, no control is exerted over metabolic fluxes: they are selected at random as long as they are permitted by stoichiometry, resulting in broad yet non-trivial flux distributions that support a small, non-zero growth rate. In the FBA limit, fluxes are controlled precisely to  maximize the growth rate, with zero fluctuations. The existence of these two limits defines a fundamental, and still unanswered, question about metabolic networks: Is there empirical evidence that real metabolic networks are located in an intermediate regime between the two limits where fluctuations are non-negligible~\cite{de2016growth}, and if so, what are the properties of this intermediate regime? Here, we address this question using metabolic flux and single-cell physiology data for \emph{Escherichia coli}.

\section{Results}
We start by considering a set of metabolic reactions in the well-mixed, continuum limit. Let $S_{i \mu}$ be the stoichiometric coefficient of the metabolite $\mu$ (whose concentration is $c_\mu$) in reaction $i$, whose flux is $v_i$. The metabolic network dynamics is then given by mass balance equations:
\begin{equation}
\dot{c}_\mu = \sum_i S_{i \mu} v_i.
\end{equation}
Assuming homeostasis (i.e., steady state, $\dot{c}_\mu=0$) and including further constraints from thermodynamics, nutrient availability, and kinetic limits in the form of lower (LB) and upper (UB) bounds on fluxes, we obtain a convex polytope $\mathcal{P}$ of feasible steady states (``metabolic phenotypes'') in the space of fluxes:
\begin{eqnarray}
\sum_i S_{i \mu} v_i = 0, \nonumber \\
v_i \in [ v_i^{\rm LB},v_i^{\rm UB} ]. \label{polytope}
\end{eqnarray}
In addition to \emph{bona fide} chemical reactions, constraint-based models often include a phenomenological ``biomass reaction'' in the form of a linear combination of metabolite fluxes, $\lambda(\vek{v})=\sum_i \xi_i v_i$, where the proportions $\xi_i$ are set to mimic cell growth, i.e., the metabolite fluxes necessary to reconstitute the biomass of a new cell in a typical  division time. 

Flux balance analysis looks for the flux configuration $\vek{v}_{\rm max}$ that maximizes growth $\lambda_{\rm max}=\lambda(\vek{v}_{\rm max})$ subject to constraints given by Eqs~(\ref{polytope}), which can be easily found by linear programming. In contrast, our maximum entropy approach starts with a distribution over fluxes with a Boltzmann form, which assumes that the fluxes are as random as possible while achieving a desired average growth rate~\cite{de2016growth}:
\begin{equation}
P_{\beta}(\bf{v}) =
\begin{cases}
 \frac{1}{Z} e^{\beta \lambda(\vek{v})} & \vek{v}\in \mathcal{P},\\
 0 & \vek{v}\notin \mathcal{P}. 
\end{cases} \label{mem}
\end{equation}
The parameter, $\beta$, of the distribution $P$ can then be set to match the predicted average growth rate to the measured growth rate, $\lambda_{\rm data}$:
\begin{equation}
\bar{\lambda}(\beta)  = \int_{\vek{v}\in\mathcal{P}} d\vek{v}\; \lambda(\vek{v}) P_\beta(\vek{v}) = \lambda_{\rm data}.
\end{equation}
Alternatively, $\beta$ could also be fit to match (e.g., in a $\chi^2$-sense) some subset of fluxes that can be  measured experimentally. Once $\beta$ is fixed, the joint distribution of Eq.~(\ref{mem}) can be queried for average fluxes, flux correlations, or other quantities of interest that we  discuss later.

\begin{figure}
\centering
\includegraphics[width=\linewidth]{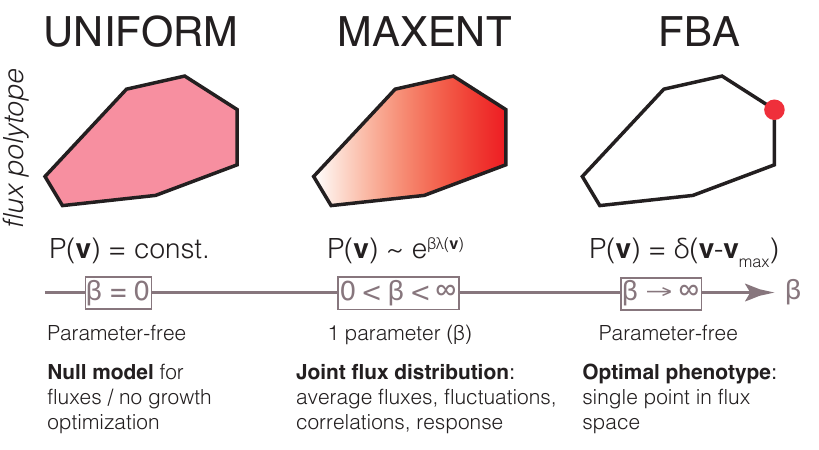}
  \caption[]
  { 
\textbf{Models of joint distribution over fluxes in a metabolic network.} 
Stoichiometric and thermodynamic constraints define a high-dimensional convex polytope of permissible fluxes, here shown in cross-section schematically  as a black polygon. In a uniform model, the flux distribution, $P(\vek{v})$, is uniform over this polytope (left). In contrast, flux balance analysis (FBA) finds a single permissible and optimal combination of fluxes, $\vek{v}_{\rm max}$ (right polytope, red dot), such that the growth rate is maximal, $\lambda_{\rm max}$. FBA and the uniform model are two limits (of $\beta\rightarrow \infty$ and $\beta = 0$, respectively) of a one-parameter family of distributions (middle polytope), where increasing the parameter $\beta$ biases the flux distribution (red gradient) away from uniform towards achieving higher average growth rates, $ \bar{\lambda}(\beta) $. The distribution over fluxes has a Boltzmann form from statistical physics and corresponds to a case where fluxes are as random as possible while achieving a specified growth rate.  
}
  \label{f1}
\end{figure}

The maximum entropy distribution with a constrained average growth rate has two interesting limits, as illustrated in~\figref{f1}. The growth rate, $\bar{\lambda}$, increases with $\beta$ (which we will refer to as an ``optimization parameter'') until, in the limit $\beta\rightarrow \infty$, the distribution $P_\infty(\vek{v})$ collapses into a delta function at $\vek{v}_{\rm max}$, lying at the boundary of the polytope $\mathcal{P}$: this is the FBA solution that supports the maximal growth rate $\lambda_{\rm max}$. Conversely, as $\beta\rightarrow 0$, Eq~(\ref{mem}) yields a uniform sampling of fluxes over the permitted polytope $\mathcal{P}$: this ``uniform'' solution is an interesting baseline case for comparison because it incorporates all stoichiometric constraints but postulates no  regulation-mediated growth rate optimization. In statistical physics, high-$\beta$ regime (limiting toward the FBA solution) corresponds to the ``energy-dominated'' regime, while the low-$\beta$ regime (limiting toward the uniform sampling) corresponds to the ``entropy-dominated'' regime; the optimization parameter $\beta$ corresponds to the inverse temperature.

\begin{figure*}
\centering
\includegraphics[width=16cm]{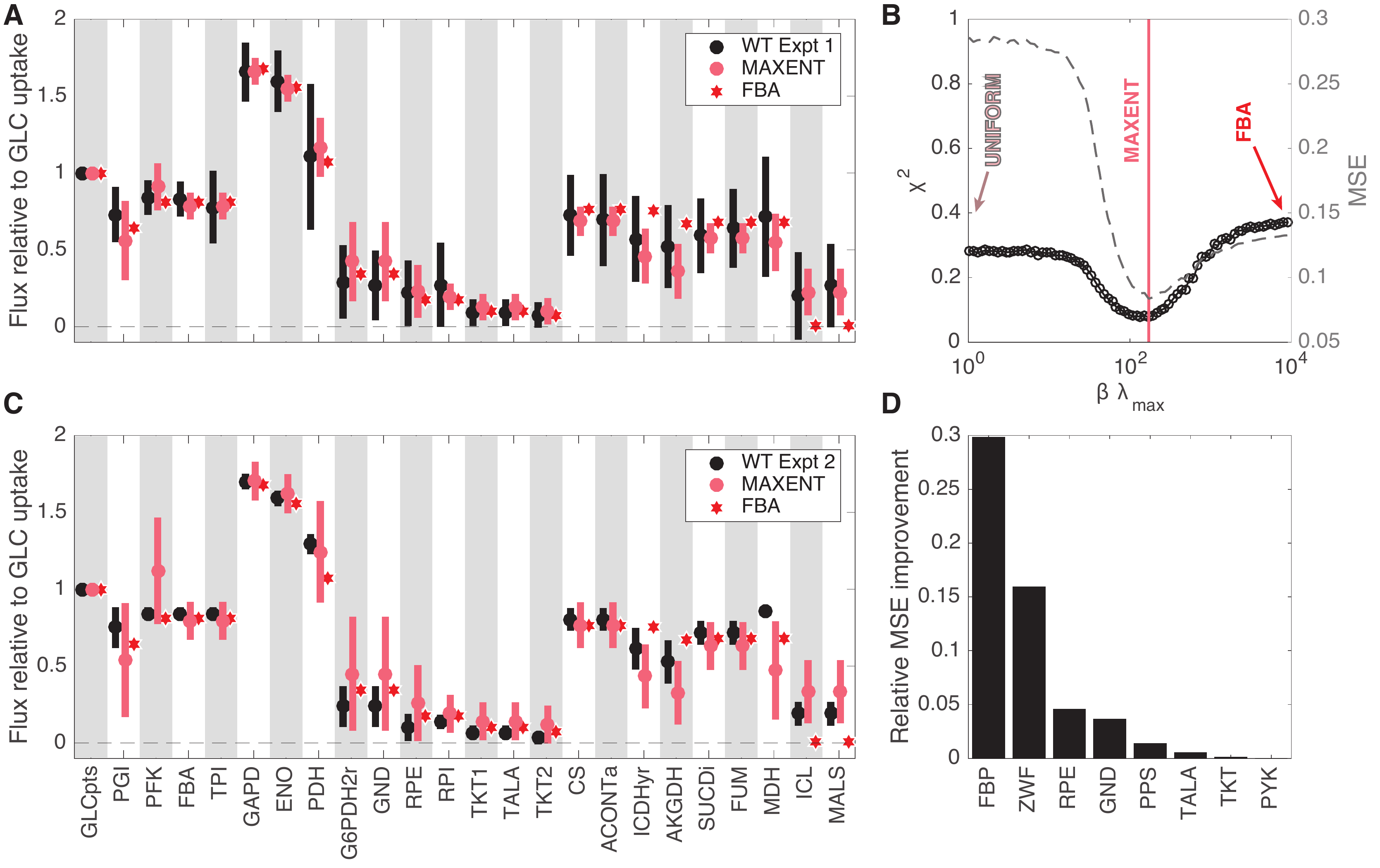}
  \caption[]
 { 
 \textbf{Maximum entropy model outperforms FBA flux predictions in \emph{Escherichia coli} during steady state growth.}
 {\bf (A, C)} Comparison of measured fluxes (black, mean $\pm$ SD over biological replicates; normalized to glucose uptake) with predictions of FBA (red stars) and of the maximum entropy model (pink, mean $\pm$ SD from the predicted joint distribution over fluxes). Data for (A) are a collection of 35 experiments from Ref~\cite{cecafdb}; data for (C) are three replicates from Ref~\cite{ishii2007multiple} . Wild type \emph{E. coli} was grown in glucose-limited medium with low dilution/growth rates (below $0.4\e{h}^{-1}$, no acetate excretion).  
 {\bf (B)} Goodness-of-fit between flux measurements in (A) and the maximum entropy predictions, as a function of dimensionless $\beta\lambda_{\rm max}$ parameter. $\chi^2$ (left axis, black) weights the (squared) error in each flux prediction by the sum of measured experimental variance and the maximum entropy predicted variance (see text); MSE (right axis, gray) is an unweighted mean squared error between measurements and predictions. In both measures, the maximum entropy model (pink line at an intermediate value of $\beta^* \lambda_{\rm max} \approx  170$) provides a better fit than the uniform ($\beta=0$) or FBA ($\beta\rightarrow\infty$) limits.
 {\bf (D)} Improvement in goodness-of-fit when using maximum entropy model over FBA for flux predictions in 7 \emph{E. coli} mutant strains deleted for metabolic enzymes indicated on the horizontal axis; data from Ref~\cite{ishii2007multiple}.
 }
 \label{f2}
\end{figure*}

\subsection{Flux predictions for \emph{Escherichia coli}}

We constructed a maximum entropy model for the catabolic core of the \emph{E. coli} metabolic network from the genome scale reconstruction iAF1260 (see SI Appendix), in a glucose-limited minimal medium in aerobic conditions~\cite{orth2011comprehensive}. The network comprises $N=86$ reactions among $M=68$ metabolites and includes glycolysis, pentose phosphate pathway, TCA cycle, oxidative phosphorylation, and nitrogen catabolism. The dimension of the resulting polytope $\mathcal{P}$ of allowed steady states is $D=23$, from which we can efficiently draw flux configurations according to Eq.~(\ref{mem}) using Hit-and-Run Monte Carlo Markov Chain after suitable preprocessing~\cite{de2015uniform} (see SI Appendix).

To evaluate the predictions of the maximum entropy model, we compared $N_f=23$ measured metabolic fluxes in \emph{Escherichia coli} from previously published data to our predictions, as shown in~\figref{f2}A,C. As a goodness-of-fit measure we defined $\chi^2 = N_f^{-1} \sum_{i=1}^{N_f} \left(\langle v_i\rangle - V_{i}\right)^2/\left(\sigma^2_i + E^2_{i}\right)$, where $V_i$ is the measured flux (relative to glucose uptake) and $E_i^2$ the associated measurement error variance, while $\langle v_i\rangle$ and $\sigma_i^2$ are the mean (and variance, respectively) of the corresponding flux computed in the maximum entropy model of Eq.~(\ref{mem}). We chose $\beta$ that minimized $\chi^2$, as shown in~\figref{f2}B, to make our flux predictions. $\chi^2$ at the optimal parameter $\beta$ was lower than in either the FBA or uniform limits, and this remained robustly true even with alternative goodness-of-fit measures (e.g., mean-squared-error, MSE=$N_f^{-1} \sum_{i=1}^{N_f} \left(\langle v_i\rangle - V_{i}\right)^2$ that  does not normalize by a variance, $\sigma^2_i$, which grows as $\beta$ is decreased). 

The maximum entropy model thus provided a better account of the measured metabolic fluxes for wild type \emph{E. coli} at intermediate values of $\beta$, rather than in the limit $\beta\rightarrow\infty$, which corresponds to the flux balance analysis (FBA) solution. Specifically, for both datasets analyzed in~\figref{f2}, the best-fitting dimensionless combination was $\beta^* \lambda_{\rm max}\sim 10^2$. In addition to a better quantitative match overall, the maximum entropy model correctly predicted non-zero flux through the glyoxylate shunt, i.e., for the ICL and MALS reactions, which FBA misses qualitatively by setting them to zero. As a consequence, this also leads to a better match of our model with data for reactions ICDHyr and AKGDH that channel pyruvate through the Krebs cycle.
 Lastly, the maximum entropy model also improved on the FBA predictions for several mutant strains of \emph{E. coli} that were deleted for various metabolic genes, as shown in~\figref{f2}D.

\subsection{Fluxes, variances, and correlations along the optimization trajectory}

\begin{figure*}[t]
\centering
\includegraphics[width=\linewidth]{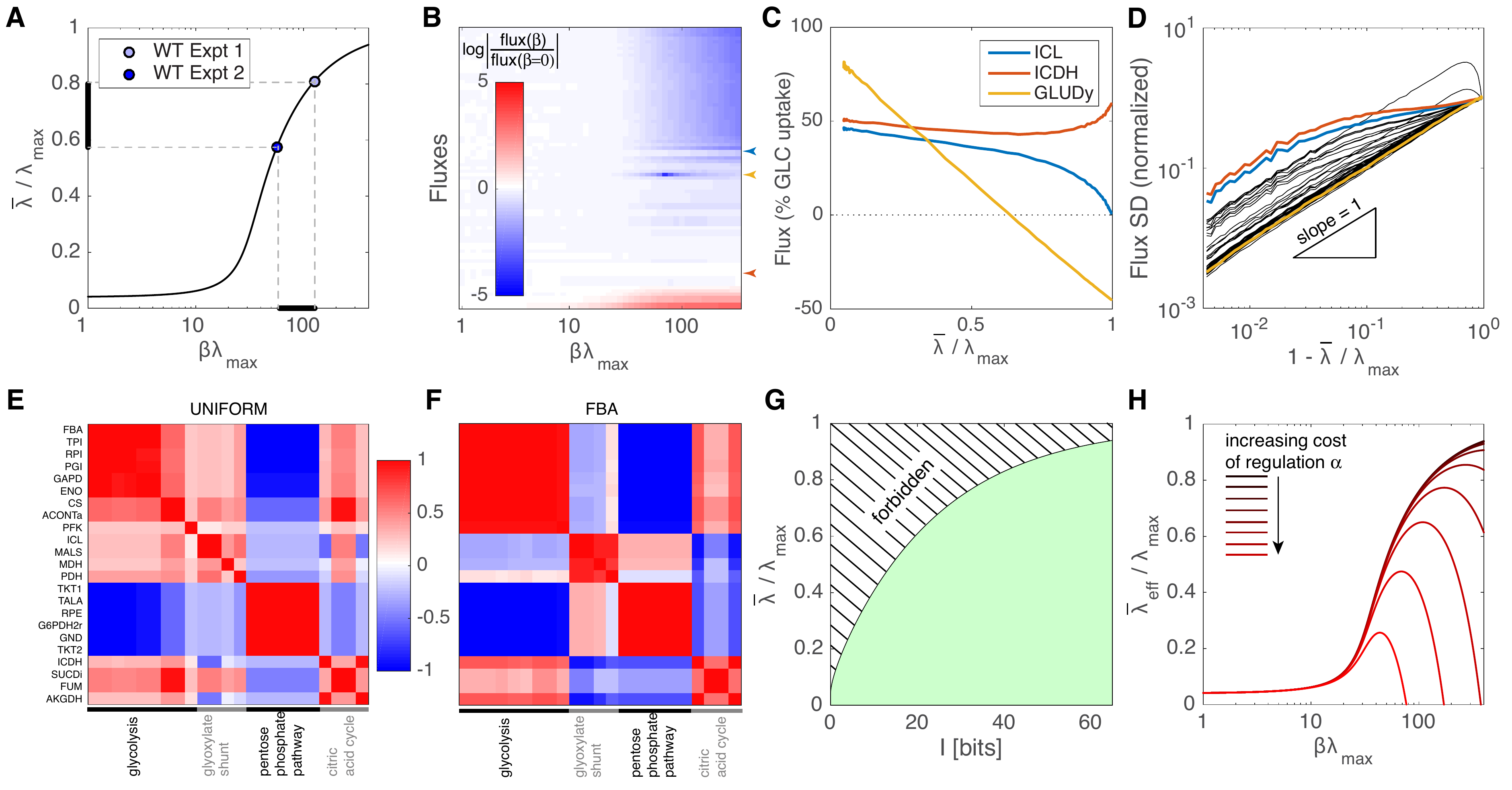}
  \caption[]
 {\textbf{Behavior of the maximum entropy model as a function of growth rate optimization parameter, $\beta\lambda_{\rm max}$.}
 {\bf (A)} Average growth rate (relative to maximal achievable rate) as a function of $\beta\lambda_{\rm max}$. $\beta^*\lambda_{\rm max}$ values that fit best the measured fluxes in~\figref{f2}A,C are shown as blue points; thick black lines on the axes denote the corresponding expected ranges for wild type \emph{E. coli} growth. 
 {\bf (B)} Log fold change of average fluxes as a function of $\beta\lambda_{\rm max}$, relative to the uniform distribution; fluxes are sorted by their change in the FBA ($\beta\rightarrow\infty$) limit. 
 {\bf (C)} Dependence of three selected fluxes (legend, highlighted with correspondingly colored arrows in (B)) on the average growth rate, $\bar{\lambda}$: ICL is turned off in the FBA limit (but not in the maximum entropy solution), ICDH remains nearly unchanged with $\bar{\lambda}$, and GLUDy is the only flux that switches sign. 
 {\bf (D)} Flux fluctuations (each black line = SD of one flux according to maximum entropy distribution, three selected fluxes from (C) highlighted in color) scale linearly with distance to maximal growth rate and vanish in the FBA limit.  
 {\bf (E, F)} Correlation coefficient matrix (Pearson correlation, color scale) between 23 selected fluxes in the uniform (E) and FBA (F) limits, computed within maximum entropy framework. Fluxes have been grouped into four clusters according to the correlation in the FBA limit and reordered accordingly in both plots; clusters are strongly enriched for fluxes belonging to pathways denoted at bottom. Note the flip in correlation sign between the glycolysis and glyoxylate shunt pathways between the two limits. 
 {\bf (G)} Achieving a particular growth rate (y-axis) requires reducing the entropy of the joint distribution of fluxes at least by $I$ bits below the entropy of the uniform distribution (green region). Points in the hashed (forbidden) region are not achievable.
 {\bf (H)} A simple model in which tight regulation of fluxes (higher $I$) enables higher growth rates, as in (G), but also entails metabolic cost (see text). For a given cost $\alpha$, the effective growth rate $\bar{\lambda}_{\rm eff}$ is maximized at an intermediate value of $\beta\lambda_{\rm max}$. 
 }
 \label{f3} 
\end{figure*}

It is instructive to examine the evolution of the joint distribution over fluxes, $P_\beta(\vek{v})$, as a function of the optimization parameter, $\beta$. \figref{f3}A shows how the growth rate approaches the maximal rate achievable, $\lambda_{\rm max}$, with the inferred values of $\beta^*$ from~\figref{f2}A,C suggesting an ``optimization level'' in the range of $\sim 60-80\%$ of the maximum. These levels are reached by adjusting flux values away from what they would have been under uniform sampling from the polytope of the allowed metabolic phenotypes, $\mathcal{P}$. \figref{f3}B traces the relative changes in all fluxes as a function of $\beta$. Interestingly, in the FBA limit almost half of the fluxes (38 out of 86 fluxes, the upper half of the plot) are forced to zero, whereas at values that fit data best ($\beta^*\lambda_{\rm max}\sim 10^2$) these fluxes only decrease by about $1/3$ relative to their average value in the uniform sampling limit. Furthermore, the glyoxylate shunt remains active, in agreement with experimental observations. Surprisingly, only for a few reactions the fluxes are predicted to increase with growth rate optimization relative to the uniform sampling (lowest $\sim 5$ fluxes in~\figref{f3}B). These are mainly nitrogen and phosphate transport reactions, and to a lesser extent, MDH and PGI reactions; the latter two reactions are classified as reversible, so the predicted increase may have thermodynamic rather than regulatory reasons. 

We separately illustrate three flux behaviors in~\figref{f3}C, for isocytrate lyase (ICL), dehydrogenase (ICDH), and glutamate dehydrogenase (GLUDy). ICL and ICDH track the relative channeling of carbon sources in the Krebs cycle vs glyoxylate shunt; ICL flux is switched off in the $\beta\rightarrow\infty$ limit, whereas ICDH flux remains nearly constant with $\beta$. In contrast, GLUDy reaction is reversible, switching sign at intermediate values of $\beta$, while at high $\beta$ the reaction ultimately gets frozen in the backward direction, implying high levels of ammonia in the cell.

We also evaluated the predicted fluctuations in metabolic fluxes from the maximum entropy model, Eq.~(\ref{mem}), at $\beta^*\lambda_{\rm max}\sim 10^2$, and found a clear division between reactions with high and low coefficients of variation ($CV$). Among tightly controlled fluxes were all glycolytic reactions ($CV<0.3$) with the exception of PGI, as well as all transport reactions related to biomass formation (i.e., for glucose, oxygen, ammonia, carbon dioxide, phosphate ions; $CV < 0.11$), the first part of the Krebs cycle, and the irreversible reactions of oxidative phosphorylation. 

We next wondered how flux variances scale with the optimization level. In the uniform sampling limit ($\beta=0$) the variances should be large, characteristic of the shape and extent of the  permitted polytope, $\mathcal{P}$. While in the FBA limit ($\beta\rightarrow\infty$) the flux fluctuations should vanish, we expect a well-defined scaling regime at high $\beta$ where the variances  shrink towards the FBA solution in a manner that is independent of the global polytope properties. This regime is indeed reached for all fluxes at $\bar{\lambda}/\lambda_{\rm max} \gtrsim 0.90$ and for some fluxes much earlier, as shown in~\figref{f3}D: flux fluctuations subsequently decrease with $\beta$  as $\sigma_i(\beta)/\sigma_i(\beta=0) \propto (1-\bar{\lambda}(\beta)/\lambda_{\rm max})$.

What kind of correlation structure between fluxes does the maximum entropy model predict? While the growth rate $\lambda$ is linear in constituent fluxes in Eq.~(\ref{mem}), suggesting that the joint distribution could factorize, correlations between fluxes develop because of the stoichiometric constraints that define the polytope $\mathcal{P}$. A subset of fluxes that we focus on in~\figref{f3} exhibits a clear structure of strong (anti-)correlation both under uniform sampling (\figref{f3}E) and in the FBA limit (\figref{f3}F). The FBA pattern of correlations, in particular, can easily be partitioned into four groups using a clustering algorithm~\cite{iclust} so that the groups are strongly enriched for reactions characteristic of glycolysis, glyoxylate shunt, pentose phosphate pathway, and citric acid cycle, respectively. Fluxes in the glycolysis cluster tend to correlate strongly with fluxes in the citric acid cycle cluster, but anti-correlate with glyoxylate shunt and pentose phosphate pathway cluster. Comparison of the FBA  correlations (F) with the uniform sampling (E) reveals that stoichiometric constraints alone shape much of the correlation structure, with the exception of anti-correlation between glycolysis and glyoxylate shunt clusters, which is a distinct consequence of the growth rate optimization. More generally, it is intriguing to apply maximum entropy to recover the correlation structure of metabolic fluxes in the FBA limit and use that to identify, automatically via clustering, separate metabolic pathways (see SI for correlation between all fluxes).

\subsection{Lower limit to regulatory information required for high growth rates}

As growth rate optimization parameter $\beta$ is increased, flux variances shrink (\figref{f3}D), correlations strengthen (\figref{f3}F), and the distribution over fluxes within the polytope $\mathcal{P}$ localizes closer to the FBA solution, $\vek{v}_{\rm max}$. The degree of localization can be quantified by the entropy of the joint distribution over fluxes:
\begin{equation}
S(\beta) = -\int_{\vek{v}\in\mathcal{P}} d\vek{v}\; P_\beta(\vek{v}) \log P_\beta(\vek{v}).
\end{equation}
Because $P_\beta(\vek{v})$ is, by construction, a \emph{maximum} entropy distribution with average growth rate $\bar{\lambda}(\beta)$, the decrease in entropy, $I(\beta) = S(\beta=0)-S(\beta)$, is a measure for the \emph{minimal} amount of information necessary to control the fluxes and achieve a given average growth rate. This is shown graphically in~\figref{f3}G, where we plot the average growth rate, $\bar{\lambda}$, as a function of information, $I$ (expressed in bits), parametrically in $\beta$. The resulting curve divides the $(I,\bar{\lambda})$ plane into two halves: while it is possible to achieve metabolic phenotypes below the $I(\bar{\lambda})$ curve, the dashed region above the curve is forbidden. This is because \emph{no} distribution exists that achieves high growth rates $\bar{\lambda}$ without also deviating from the uniform distribution by at least the required number of bits.

\figref{f3}G suggests that at least $\sim 40$ bits of information are required to control the fluxes and reach growth rates amounting to $\sim 80\%$ of the maximal rate, $\lambda_{\rm max}$, as reported in data for \emph{E. coli}; higher growth rates call for increasing amounts of information, which formally diverges in the FBA-limit as $\beta\rightarrow\infty$. 

Cells control metabolic fluxes through regulatory networks, either indirectly, by regulating the expression of metabolic enzymes, or directly, by modulating the enzymatic activity through various feedback loops; either way, metabolic resources are required to exert this control. This leads to a trade-off: flux control is necessary to support a high growth rate, but itself carries a growth rate penalty. We created a simple toy model to capture this intuition (see SI Apprndix). Here, $K$ regulatory pathways control the fluxes and each pathway is modelled as a Gaussian information channel, so that together, these channels provide $I(\beta)$ bits of necessary information as shown in~\figref{f3}G. The signal-to-noise of each regulatory channel is determined by the number of regulatory molecules: higher molecular counts enable precise control and thus higher information, but impose higher cost. In this model, the cost-free growth rate at given $\beta$ is reduced by the cost to support $K$ channels which control the fluxes, so that the resulting effective growth rate is:
\begin{equation}
\bar{\lambda}_{\rm eff}(\beta) = \bar{\lambda}(\beta) - \alpha K \left[ 2^{2I(\beta)/K} - 1\right],
\end{equation}
where $\alpha$ determines the metabolic cost of regulatory molecules, and we estimated the number of regulatory pathways, $K$, to be approximately the number of degrees of freedom of the flux polytope, $K\approx D$. The cost of regulation clearly limits the achievable growth rate, as shown in~\figref{f3}H, where the $\bar{\lambda}_{\rm eff}(\beta)$ curves now develop a maximum rather than increasing monotonically as in the cost-free case of~\figref{f3}A. While our toy model is very simplistic, it does capture properly the scaling of information with the growth rate, as well as the exponential metabolic cost of achieving high information transmission in molecular networks, reported previously~\cite{gtawreview,annurevgt}. Thus, among many possible constraints acting on a cell, the cost of regulating metabolism itself \cite{kacser1995control} can impose non-negligible limits to growth.

\subsection{Experimental test of growth rate fluctuation scaling}
\begin{figure}
\centering
\includegraphics[width=\linewidth]{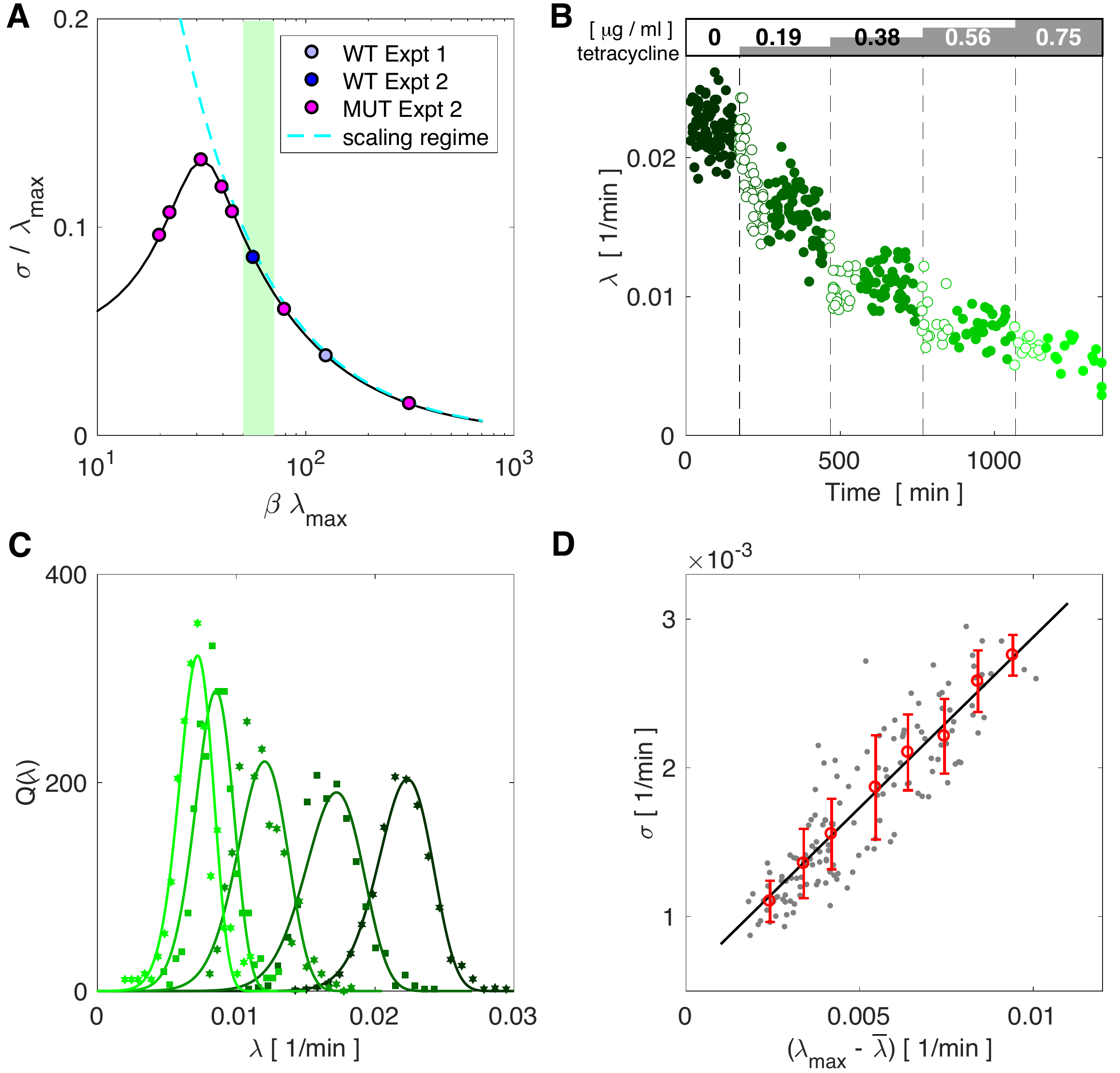}
  \caption[]
 { \textbf{Maximum entropy model predicts growth rate fluctuations across individual cells in an isogenic population.}
 {\bf (A)} Predicted fluctuations in the growth flux, $\sigma/\lambda_{\rm max}$, as a function of $\beta\lambda_{\rm max}$. Two wild type experiments from~\figref{f2}A,C are shown as blue points, mutant flux phenotypes from~\figref{f2}D shown as magenta points, approximate range experimentally probed in~\figref{f4}B-D shown as green shade.  In the scaling regime (cyan dashed line), $\sigma \propto (\beta\lambda_{\rm max})^{-1}$. 
 {\bf (B)} Single-cell measurements (each dot = one division event) of elongation rate in a microfluidic ``mother machine'' device for wild type \emph{E. coli} under increasing sub-inhibitory antibiotic concentrations, indicated on top; data from a single experiment in Ref~\cite{tobianna}. After switching to higher antibiotic concentration (lighter shade of green), cells show transient behavior (empty green circles) which we ignore and focus only on steady state (full green circles).
 {\bf (C)} Measured distributions (plot symbols) of growth rates from multiple experiments as in (B) (different shades of green = different tetracycline concentration). Solid lines show maximum entropy distributions with best-fit values for $(\beta,\lambda_{\rm max})$ (SI Appendix).
 {\bf (D)} Testing the predicted scaling of growth rate fluctuations. Each dot = one lineage from experiments in \figref{f4}C.  The average growth rate, $\bar{\lambda}$, the maximal observed growth rate, $\lambda_{\rm max}$, and the SD of the growth rates, $\sigma$, were estimated separately on each lineage. Scaling regime of \figref{f4}A predicts a linear relationship, which is shown here by binned data (red points with errorbars = mean $\pm$ SD within equi-distant bins) and best linear fit (solid black line, $R^2=0.77$).
 }
 \label{f4} 
\end{figure}

Can we test the novel predictions of our theory that extend beyond the domain of validity of the flux balance analysis? While it is experimentally unfeasible to measure metabolic fluxes and their fluctuations at the single cell level, one can tractably measure division times and growth rates for single \emph{Escherichia coli} cells growing in stable conditions for long periods of time. In our model, such growth measurements directly connect to the biomass producing reaction with its associated ``growth flux,'' $\lambda(\vek{v})$. \figref{f3}D suggests that flux fluctuations should scale $\propto (\lambda_{\rm max}-\bar{\lambda})$, and since the growth flux is a linear combination of \emph{bona fide} metabolic fluxes, its fluctuations, too, should follow the same scaling. To verify this explicitly, we computed the fluctuations in growth rate, $\sigma/\lambda_{\rm max}$, as a function of the optimization parameter $\beta$, in~\figref{f4}A. In the range of $\beta\lambda_{\rm max}\gtrsim 40$, characteristic of both wild type \emph{E. coli} experiments and many mutants, the predicted growth fluctuations indeed obey
\begin{equation}
\frac{\sigma}{\lambda_{\rm max}} \propto (\beta\lambda_{\rm max})^{-1} \propto  (\lambda_{\rm max}-\bar{\lambda}); \label{scaling1}
\end{equation}
we refer to this range as the ``scaling regime.'' Beyond variance, the complete distribution of growth rates, $Q(\lambda)$, can be sampled by marginalizing the maximum entropy model, Eq.~(\ref{mem}).

Measurements of single-cell growth rates allow us to estimate growth rate distributions and compare them to the predicted $Q(\lambda)$, as well as to empirically extracted $\bar{\lambda}$, $\lambda_{\rm max}$, and the fluctuations $\sigma$, to verify the predicted relation of Eq.~(\ref{scaling1}). We used previously published data~\cite{tobianna} where \emph{E. coli} cells were stably grown in the ``mother machine'' microfluidic device while multiple sub-inhibitory steps of concentration  of the antibiotic tetracycline were delivered as shown in~\figref{f4}B. Low concentrations of antibiotic allowed us to probe different average growth rates in the same setup, and to construct empirical distributions of growth rates for every antibiotic concentration by pooling data from technical replicates of the multi-step experiments (SI Appendix). We find an excellent match between measured and predicted growth rate distributions in~\figref{f4}C for all five concentrations of the antibiotic used. Looking at many individual lineages in separate microfluidic channels, we can also extract $\lambda_{\rm max}$, $\bar{\lambda}$, and $\sigma$ per lineage empirically and confirm the predicted scaling of growth rate fluctuations, as shown in~\figref{f4}D.

\subsection{Fluctuation-response relationship and  associated timescales}
Our approach so far has been limited to modeling steady state growth.  Under mild conditions, we can also study the dynamical response of the network in linear regime under small perturbations. 
A simple biologically-motivated dynamics is provided by the minimal model of diffusion-replication inside the metabolic space~\cite{de2016growth}, described by the one-parameter ($D$, the diffusion coefficient) equation for the growth rate distribution $p(\la)$:  
\begin{equation}
\dot{p}(\lambda) = (\la-\overline{\la})
p(\la) +D \left[\frac{\partial^2 p}{\partial\la^2} -\frac{\partial}{\partial\la} \left[p(\la)
\frac{\partial}{\partial\lambda}(\log q(\la))\right]\right]~~, \nonumber
\end{equation} 
where $q(\la)$ is the marginal distribution of the growth rate at uniform sampling ($\beta=0$). An analytical solution of this equation can be obtained in the limit of small $D$, leading to the following scaling laws for the typical response times as well as the growth rate fluctuation autocorrelation time $\tau$~\cite{demaso}:
\begin{align*}
& \sigma \sim \tau^{-1} \sim \la_{\rm max}-\overline{\la}\sim D^{1/3}. 
\end{align*}
These experimentally testable relations predict a divergent slowing down of the response time with growth rate maximization. 
As a consequence, growth rate fluctuations could take on a  functional role in speeding up the response to environmental perturbations, e.g., to nutritional up-shifts or externally applied stresses.  Even if  the experimental test of such dynamic predictions is beyond the scope of this paper, our model  connects to a wide range of currently ongoing metabolism- and growth-related  investigations.

\section*{Discussion}
In this work we considered maximum entropy distributions at fixed average growth rate in the space of metabolic phenotypes, a straightforward and statistically rigorous extension of the flux balance analysis, which is recovered in the asymptotic limit. Experimental estimates of enzymatic fluxes of the central carbon core metabolism in bulk cultures of \emph{E. coli}, as well as empirical growth rate distributions of \emph{E. coli} collected from single cell measurements, are consistent with intermediate level of growth optimization ($\beta\lambda_{\rm max}\sim10^2$ and $\bar{\lambda}/\lambda_{\rm max}\sim 0.6-0.8$).  We find that fluctuations can be  captured by a simple maximum entropy model, and that the zero-fluctuation FBA limit qualitatively misses  important  experimental facts, e.g., the observed non-zero fluxes through the glyoxylate shunt. 

The improved ability of our model to match the flux measurements is a consequence of a single extra parameter, $\beta$, which can easily be determined from existing experimental data. Beyond a better fit, however, our  model also makes a wide range of  predictions, extending the domain of metabolic network analysis to the single-cell level. While it is difficult to measure the single-cell metabolic fluxes and their fluctuations in isogenic populations in steady state, such measurements for the growth rate are increasingly available. This connection enables the new predictions of our theory to be tested, and opens up the theory for  verifiable extensions. Validating the predicted scaling of growth rate fluctuations in \figref{f4} is only the first step, with two broad lines of investigation within reach.

First, our approach is not limited to the core catabolism analyzed here or to bacterial metabolism, but can in principle be extended to other genome-scale networks. In practice, however, we often lack suitable large-scale experimental flux measurements. It is also likely that physico-chemical constraints alone are insufficient to yield quantitatively accurate predictions. Similar issues arise also in the core catabolism for high growth rates that exceed the threshold of the acetate switch, for which additional constraints have to be added in FBA-based approaches~\cite{mori2016constrained}. Our method can be extended to accommodate such cases, or systems where strict growth maximization is likely not a suitable objective. Extra objectives or constraints in the maximum entropy would appear as additional terms in the exponent of Eq.~(\ref{mem}), where their corresponding parameters would control various trade-offs between the objectives. This flexibility may be required to model metabolic dependencies, cell type heterogeneity, or interactions between cells.

Second, we sketched how the maximum entropy model could be extended to dynamics through fluctuation-response relations. This requires further assumptions that need to be tested separately, but makes a very strong prediction about the link between the autocorrelation time of growth fluctuations and the typical response time to, e.g., nutrient shifts. This link appears fundamental, since the response time is a central biological quantity measurable in bulk, while the fluctuation autocorrelations are microscopic, single-cell properties, which can be measured with recent experimental setups. Interestingly, the predicted response times lengthen with the degree of growth rate optimization, suggesting a trade-off between responsiveness to changes and efficiency in steady state; as a consequence, it is unclear whether the evolutionarily optimal outcome should be equated to complete growth rate optimization with no fluctuations, e.g., the FBA limit. Quantitatively, in stable environments where \emph{E. coli} grows well and possibly achieves a high degree of growth rate optimization, one could experimentally look for signatures of long-timescale fluctuations, either directly in the growth signal, or by proxy through constitutive gene expression. Curiously, we report that the parameter $\beta$ of our model has the dimension of time, whose best-fit value inferred from \emph{E. coli} data is of the order of 1 day. A proper investigation into the role of $\beta$ deserves further theoretical work.

Beyond extensions to dynamics, our analysis made two further theoretical contributions. First, it clarified the relative roles of stoichiometric constraints and the growth optimization assumption in flux balance analysis. The maximum entropy model is an explicit construction of a smooth interpolation between the uniform regime (where only stoichiometric constraints are active) and the FBA (where growth is maximized in addition). The uniform limit is a natural baseline---where no control is exerted by the cell---against which to compare the observed fluxes, their fluctuations, and correlations, as we have done in~\figref{f3}. Without this baseline comparison, it is hard to assess how surprising the observations of metabolic optimality should really be~\cite{ibarra2002escherichia}. Our second theoretical contribution is the observation that a certain minimal information is needed to achieve a desired growth rate (\figref{f3}G, H). This information is expressed in the same currency (bits) in which we measure the performance of regulatory networks, enabling us to suggest a tradeoff that sets the optimal degree of metabolic control. Contrary to other cellular networks where estimation of information only has been done for single network components or simple pathways~\cite{annurevgt}, the metabolic network is the sole case where we could estimate the lower bound on the required number of regulatory bits. Our statistical mechanics approach thus opens a connection between metabolic networks and their regulatory counterparts, which is both of theoretical interest and could also be probed in comparative genomic studies.

\begin{acknowledgments}
We acknowledge the support of the Austrian Science Fund grant FWF P28844 (G.T.) and  of the People Programme (Marie Curie Actions) of the European Union's Seventh Framework Programme (FP7/2007-2013) under REA grant agreement no. $[291734]$ (D.D.M)
\end{acknowledgments}

\appendix

\section*{Appendix}
\subsection*{The sampling Monte Carlo Algorithm}
We sampled the space of steady states of the model using a hit-and-run Markov Chain Monte Carlo algorithm, subject to  an ellipsoidal preprocessing step in order to tackle ill-conditioning. 
The uniform sampling of convex bodies by means of Monte Carlo methods has been considered a breakthrough in computational convex analysis, e.g., it makes feasible the calculation of the volume
which is otherwise a computationally difficult problem \cite{dyer1988complexity}.  A fast and  popular algorithm to sample points inside convex bodies is the hit-and-run (HR) Markov Chain Monte Carlo \cite{Turcin:1971, Smith:1996p4127}, which works along the following lines.
Given a $D$-dimensional convex  polytope $\mathcal{P}$, from which one wants to sample, and a point inside the polytope, $x_k \in \mathcal{P}$:
\begin{itemize}
\item Choose a uniformly distributed direction $\theta_k$, that is, a point generated from the uniform distribution on the $D$-dimensional unit sphere. This can be done with the Marsaglia method, i.e., by generating $D$ independent gaussian random variables with zero mean and unit variance, and then normalizing the vector to unit length.
\item
Generate $t$ uniformly on the interval $[ t_{\rm min},
t_{\rm max}  ] $, where $t_{\rm min}$ ($t_{\rm max}$) is the minimum (maximum) value of $t$ such that $x_k+t\theta_k \in \mathcal{P}$;
\item    
    Update $x_{k+1}=x_k+t\theta_k$, start again.
\end{itemize}
The starting point can be found, for instance, by interpolating between two vertices obtained by linear programming. The second step requires finding the intersections among a line and $\mathcal{P}$. In order to perform the HR dynamics we should always use a full-dimensional representation of the convex set. 
The mixing time of the HR, i.e., the time to converge to the desired distribution, scales as a polynomial of the dimensions of the body but the method can suffer from ill-conditioning; more precisely, the mixing time $\tau$ scales like \cite{Lovasz:1999p4121} 
\begin{equation}
\tau \sim O  \left(D^2 \frac{R^2}{r^2}\right),
\end{equation}
where $R,r$ are the radii of respectively the minimum inscribing and the maximum inscribed balls.
The prefactor $R/r$ can be reduced to a polynomial of $D$ by extracting $\theta_k$ from the surface of a matching ellipsoid instead of the unit sphere. The ellipsoid of maximum volume inscribed inside $\mathcal{P}$, i.e., the Loewner-John ellipsoid \cite{ball1997elementary}, would reduce the prefactor 
to $O(D)$, but the problem of finding it is NP-hard. 
Below we describe a method due to L.~Lovasz \cite{lovasz1987algorithmic} that finds in polynomial time an approximate matching ellipsoid that reduces the prefactor to $O(D^{3/2})$.
\subsubsection*{Preprocessing: the Lovasz algorithm}
We want to construct a couple of concentric ellipsoids $E,E'$ matching the polytope $\mathcal{P}$, i.e., such that $E' \subseteq \mathcal{P} \subseteq E$, where $E'$ is obtained from E shrinking by a factor $O(D^{3/2})$. This is called weak Loewner–John pair. We define a series of enclosing ellipsoids $E_k$, starting with $E$ as the sphere with center in the origin and radius $R$ large enough in order to inscribe the body, according to the following lines:
\begin{itemize}

\item    INPUT: An ellipsoid $E_k$ with its center $x_k$.
\item    Check if $x_k \in \mathcal{P}$, if yes go to 2, if no go to 1.
\item    1) Consider a hyperplane separating $x_k$ and $\mathcal{P}$, and the halfspace H enclosing $\mathcal{P}$, calculate the ellipsoid of minimal volume enclosing $H \bigcap E_k$; go to OUTPUT 1.
\item    2) Determine the endpoints of the axis of $E_k$, shrink the ellipsoid  and check if the shrunk ellipsoid $E'_k$ is inside the body. If yes, go to OUTPUT 2, if no go to 3.
\item 3) Consider the endpoint of an axis of the shrunk ellipsoid outside $\mathcal{P}$, e.g. $x'_k$; consider a hyperplane  separating $x'_k$ and $\mathcal{P}$, and the halfspace  enclosing $\mathcal{P}$; calculate the ellipsoid of minimal volume enclosing 
$H \bigcap E'_k$; go to OUTPUT 1.
\item    OUTPUT 1: A new ellipsoid $E_{k+1}$ of lower volume with center $x_{k+1}$, update $k$, repeat from INPUT.
\item    OUTPUT 2: A weak Loewner-John ellipsoid.
\end{itemize}

Upon calculating the reduction in volume of the enclosing ellipsoid after one step, it can be demonstrated that this series converges in polynomial time to a weak Loewner–John pair. We refer to \cite{bland1981ellipsoid} for demonstrations and formulae.

\subsubsection*{Source Code}
We have provided in {\tt doi:10.15479/AT:ISTA:62} a C++ code implementing the Lovasz preprocessing as well as the Hit-and-Run algorithm and the polytope representation of the metabolic network used in this study. Please refer to the  {\tt README.txt} file for further information.

\subsection*{Metabolic network and flux data}
We considered the model of the \emph{E. coli} catabolic core from the metabolic genome scale reconstruction iAF1260 \cite{orth2011comprehensive}. Apart from mass-balance (purely stoichiometric) constraints, we have considered the bounds on reaction fluxes from the reversibility assignments provided within the model and default bounds on uptakes reflecting glucose limited aerobic conditions.    
We considered the dataset from \cite{ishii2007multiple} that includes extensive flux, enzyme, mRNA and metabolite level measurements of \emph{E. coli} populations growing in steady-state in a glucose limited minimal medium in several conditions (dilution rates and/or gene knock outs).
We considered further experimental estimates of the core carbon metabolism from the database \cite{zhang2014cecafdb}, where
we collected experiments \cite{farmer1997reduction, siddiquee2004metabolic, schaub2008metabolic, peng2004metabolic, emmerling2002metabolic, hua2003responses, nanchen2008cyclic, nikel2009metabolic, li2006effect, toya2007direct, zhu2003metabolic, zamboni200913c}
 on wild type \emph{E. coli} performed in  glucose limited medium and dilution rates not exceeding $0.4$h$^{-1}$, i.e. below the acetate switch \cite{basan2015overflow}

\subsection*{Data analysis details}


\subsubsection*{Figure 2.}
Flux intensities have been normalized with respect to the glucose uptake (i.e. expressed in relative values) in order to compare datasets with different  growth rate/dilution (in all cases case below $0.4$h$^{-1}$). While the Lagrange parameter $\beta$ can be used to fix the average growth rate, in this case we consider the best fit for $\beta$ upon comparing relative flux values, i.e., we look for the minimum of  $\chi^2 = N_f^{-1} \sum_{i=1}^{N_f} \left(\langle v_i\rangle - V_{i}\right)^2/\left(\sigma^2_i + E^2_{i}\right)$, where $V_i$ is the measured flux  and $E_i^2$ the associated measurement error variance, while $\langle v_i\rangle$ and $\sigma_i^2$ are the mean (and variance, respectively) of the corresponding flux computed in the maximum entropy model. 
\begin{itemize}
\item {\em Figure 2B}\\
We considered  $10^2$ values of $\beta$ spanning the interval $\left[10^{-2},10^4\right]$ uniformly in logarithmic scale. For  each value of $\beta$ we sampled $10^5$ flux configurations according to the max entropy distribution with the previously described hit-and-run Monte Carlo algorithm from which we calculate the estimates of the averages and variances. These are then compared to experimental values through the $\chi^2$, as well as the mean square error, $MSE=N_f^{-1} \sum_{i=1}^{N_f} \left(\langle v_i\rangle - V_{i}\right)^2$.
\item {\em Figure 2A,C}\\
The flux values reported for the max entropy model refer to the value of  $\beta$ that minimizes the $\chi^2$.
\item {\em Figure 2D} \\
We simulated an enzyme knock out by removing the corresponding reaction from the model (one equality constraint). This leads to a new metabolic space that has been analyzed along the same lines  as the unperturbed one. Flux estimates for knockouts from \cite{ishii2007multiple} consist of one  single experiment with no repeats. The MSE improvement is the fractional difference between the minimum value and the value retrieved by FBA. 

\end{itemize}

\subsubsection*{Figure 3.} 
In these plots, flux intensities have been normalized with respect to the glucose uptake, while the growth rate has been rescaled by its maximum.
\begin{itemize}
\item {\em Figure 3A,G}\\
The curves were calculated analytically along the following lines. We fit the marginal probability density of the growth rate, otained numerically by Monte Carlo sampling, with a Beta distribution, i.e. in relative units $x=\lambda/\lambda_{\rm max}$:
\begin{equation}
q(x) = (a+1)(1-x)^a,
\end{equation}
finding simply $a=D-1=22$, where $D$ is the dimension of the polytope, i.e. the growth rate is maximized in a subspace of dimension $0$ (a vertex)~\cite{demaso}.
The average growth rate as a function of $\beta$ can be calculated from the normalizing factor of the max entropy distribution,
\begin{eqnarray}
Z(\beta) =\int_0^1 q(x) e^{\beta x} dx = \\
=\frac{\gamma(a+1,\beta)e^\beta (a+1)!}{\beta^{a+1}},
\end{eqnarray} 
where $\gamma$ is the lower incomplete gamma function and finally (Figure 3A) 
\begin{equation}
\overline{x} = \frac{d \log Z}{d\beta}.
\end{equation}

The minimal entropy reduction of the metabolic space upon fixing the average  growth rate (Figure 3G) has been  obtained by inverting the previous equation, $\beta=\beta(\overline{x})$, and performing a Legendre transform~\cite{de2016growth}
\begin{equation}
I(\overline{x}) \log 2 = \Delta S(\overline{x}) = \overline{x}\beta(\overline{x}) -\log Z( \beta(\overline{x})).
\end{equation}
Both curves can be calculated numerically, the first directly, upon sampling the average growth rate at fixed $\beta$, while the second can be obtained in parametric form upon integration; in both cases we have found an excellent agreement with the analytical approximation.

\item {\em Figure 3B-F}\\
Averages, variances and correlations among fluxes have been calculated numerically by sampling max entropy distribution of the metabolic space with the previously described Monte Carlo method. 
We considered  $10^2$ values of $\beta$ spanning the interval $\left[10^{-2},10^4\right]$ uniformly in logarithmic scale. For  each value of $\beta$ we sampled $10^5$ flux configurations. Figure 3F has been obtained by numerical asymptotic analysis of flux correlations in the limit $\beta \to \infty$.

\item {\em Figure 3H} \\
We considered a simple model where $K$ regulatory channels together provide the information, $I(\bar{\lambda})$, required to support a given average growth rate, $\bar{\lambda}$. Each channel is modeled as an additive Gaussian information channel whose capacity is given by $I_1 = \frac{1}{2}\log_2(1+SNR)$, where $SNR$ is the ratio of the signal to noise variance. To provide sufficient regulatory information we must have $I(\bar{\lambda}) \approx KI_1$. We assume that the variance of the signal is proportional to its mean, $\bar{S}$, so that $SNR = \alpha_1 \bar{S}$, for some constant $\alpha_1$; this is approximately true in the simplest models of biochemical reaction pathways where the dominant source of noise is the shot (Poisson) noise due to finite number of signaling molecules. We further assume that the metabolic cost of regulation per channel is also proportional to the number of signaling molecules used, i.e., to the mean signal in the channel, and thus the total cost (written as the decrease in the growth rate) is $\Delta\lambda = -\alpha_2 K \bar{S}$. We can eliminate $\bar{S}$ by writing it out in terms of $SNR$, which can in turn be expressed in terms of $I(\bar{\lambda})$, to yield an expression for the effective growth rate that takes into account the cost of regulation:
\begin{equation}
\lambda_{\rm eff}(\beta) = \bar{\lambda}(\beta) - \alpha K \left(2^{2I(\beta)/K}-1\right),
\end{equation}
where $\alpha = \alpha_2/\alpha_1$ is a constant proportional to the metabolic cost of signaling molecules.

\end{itemize}

\begin{table*}
\centering
\begin{tabular}{ | c  | c | c  |  c | c  | c | }  
\hline 
Step &  sample size & $\overline{\lambda}$   [min$^{-1}$] & $\lambda_{e,\rm max}$ [min$^{-1}$] & $\lambda_{\rm max}$ [min$^{-1}$] &  $\beta \lambda_{\max} $ [adim.]   \\ 
\hline 
$1$ &  $5204$ & $0.022$ & $0.029$ & $0.034$ & $66$ \\
$2$ & $3300$  & $0.017$ & $0.024$ & $0.029$ & $54$ \\
$3$ & $2252$  & $0.012$ & $0.018$ & $0.023$ & $45$ \\
$4$ &  $1464$ & $0.0085$ & $ 0.013 $ & $0.017$ & $45$ \\
$5$ &  $1536$ & $0.007$ & $ 0.011 $ & $0.015$ & $42$ \\
\hline
\end{tabular}
\caption{\label{tab} Average growth rate, empirical maximum, inferred maximum and level of optimization, $\beta \lambda_{\rm max}$, from growth rate data.}
\end{table*}

\subsubsection*{Figure 4.}
The growth rate data have been obtained by analyzing images from a microfluidics device (mother machine) setting with $30$ channels each containing $3$ \emph{E. coli} cells growing in steady-state in glucose-limited conditions. Their growth is monitored under increasing exposure to antibiotics (five concentration steps of tetracycline) specific for ribosome translation inhibition  for $5$h. Experiments have been performed on $3$ strains; $3$ technical repeats were carried out. Strains differed only by the fluorescent tags whose effects are not expected to have major contributions to the growth rates. 
\begin{itemize}
\item {\em Figure 4A} \\
The curve has been calculated analytically as described for Figure 3.
\item {\em Figure 4C}\\
The distributions have been obtained as follows.
We disregard measurements during the first $15$ time frames from each antibiotic step in order to obtain approximately stationary distributions; we also rejected outliers exceeding $4\sigma$. For each antibiotic step, we lump together growth rate measurements pertaining to different channels, time, cell order in the channel (mother-daugthers), strains and repeats; this results in a sample size of $\simeq 1-5 \cdot 10^3$ per antibiotic step.  We performed a best fit ($\chi^2$ minimization) of max entropy marginal growth rate distributions in the analytical approximation outlined in discussion for Figure 3 (above), and we retrieved the values for the two parameters $\beta$ and $\lambda_{\rm max}$ that we report in Table~\ref{tab}, alongside of the average and empirical maximum of the growth rate.

\item {\em Figure 4D}:\\ 
Each data point stands for a distribution obtained upon lumping growth rate data measured in different channels ($30$) and time points ($\simeq 3$), i.e. it constitutes a unique sample whose size is $O(10^2)$.
Each data point refers thus to a different antibiotic step ($5$), to  
a different cell ordering in the channel ($3$), strain ($3$) and repeat ($3$) for a total of $135$ points.  
We considered the usual estimators for the average and variance, while we have reported in all cases the empirical maximum for the growth rate.
\end{itemize}

\bibliography{referenceall}

\providecommand{\noopsort}[1]{}\providecommand{\singleletter}[1]{#1}
\begin{thebibliography}{55}%
\makeatletter
\providecommand \@ifxundefined [1]{%
 \@ifx{#1\undefined}
}%
\providecommand \@ifnum [1]{%
 \ifnum #1\expandafter \@firstoftwo
 \else \expandafter \@secondoftwo
 \fi
}%
\providecommand \@ifx [1]{%
 \ifx #1\expandafter \@firstoftwo
 \else \expandafter \@secondoftwo
 \fi
}%
\providecommand \natexlab [1]{#1}%
\providecommand \enquote  [1]{``#1''}%
\providecommand \bibnamefont  [1]{#1}%
\providecommand \bibfnamefont [1]{#1}%
\providecommand \citenamefont [1]{#1}%
\providecommand \href@noop [0]{\@secondoftwo}%
\providecommand \href [0]{\begingroup \@sanitize@url \@href}%
\providecommand \@href[1]{\@@startlink{#1}\@@href}%
\providecommand \@@href[1]{\endgroup#1\@@endlink}%
\providecommand \@sanitize@url [0]{\catcode `\\12\catcode `\$12\catcode
  `\&12\catcode `\#12\catcode `\^12\catcode `\_12\catcode `\%12\relax}%
\providecommand \@@startlink[1]{}%
\providecommand \@@endlink[0]{}%
\providecommand \url  [0]{\begingroup\@sanitize@url \@url }%
\providecommand \@url [1]{\endgroup\@href {#1}{\urlprefix }}%
\providecommand \urlprefix  [0]{URL }%
\providecommand \Eprint [0]{\href }%
\providecommand \doibase [0]{http://dx.doi.org/}%
\providecommand \selectlanguage [0]{\@gobble}%
\providecommand \bibinfo  [0]{\@secondoftwo}%
\providecommand \bibfield  [0]{\@secondoftwo}%
\providecommand \translation [1]{[#1]}%
\providecommand \BibitemOpen [0]{}%
\providecommand \bibitemStop [0]{}%
\providecommand \bibitemNoStop [0]{.\EOS\space}%
\providecommand \EOS [0]{\spacefactor3000\relax}%
\providecommand \BibitemShut  [1]{\csname bibitem#1\endcsname}%
\let\auto@bib@innerbib\@empty
\bibitem [{\citenamefont {Kacser}\ \emph {et~al.}(1995)\citenamefont {Kacser},
  \citenamefont {Burns},\ and\ \citenamefont {Fell}}]{kacser1995control}%
  \BibitemOpen
  \bibfield  {author} {\bibinfo {author} {\bibfnamefont {H.}~\bibnamefont
  {Kacser}}, \bibinfo {author} {\bibfnamefont {J.~A.}\ \bibnamefont {Burns}}, \
  and\ \bibinfo {author} {\bibfnamefont {D.~A.}\ \bibnamefont {Fell}},\
  }\href@noop {} {\enquote {\bibinfo {title} {The control of flux},}\ }
  (\bibinfo {year} {1995})\BibitemShut {NoStop}%
\bibitem [{\citenamefont {Orth}\ \emph {et~al.}(2010)\citenamefont {Orth},
  \citenamefont {Thiele},\ and\ \citenamefont {Palsson}}]{Orth:2010if}%
  \BibitemOpen
  \bibfield  {author} {\bibinfo {author} {\bibfnamefont {J.}~\bibnamefont
  {Orth}}, \bibinfo {author} {\bibfnamefont {I.}~\bibnamefont {Thiele}}, \ and\
  \bibinfo {author} {\bibfnamefont {B.~O.}\ \bibnamefont {Palsson}},\ }\href
  {\doibase 10.1038/nbt.1614} {\bibfield  {journal} {\bibinfo  {journal}
  {Nature Biotechnology}\ }\textbf {\bibinfo {volume} {28}},\ \bibinfo {pages}
  {245} (\bibinfo {year} {2010})}\BibitemShut {NoStop}%
\bibitem [{\citenamefont {Ibarra}\ \emph {et~al.}(2002)\citenamefont {Ibarra},
  \citenamefont {Edwards},\ and\ \citenamefont
  {Palsson}}]{ibarra2002escherichia}%
  \BibitemOpen
  \bibfield  {author} {\bibinfo {author} {\bibfnamefont {R.~U.}\ \bibnamefont
  {Ibarra}}, \bibinfo {author} {\bibfnamefont {J.~S.}\ \bibnamefont {Edwards}},
  \ and\ \bibinfo {author} {\bibfnamefont {B.~O.}\ \bibnamefont {Palsson}},\
  }\href@noop {} {\bibfield  {journal} {\bibinfo  {journal} {Nature}\ }\textbf
  {\bibinfo {volume} {420}},\ \bibinfo {pages} {186} (\bibinfo {year}
  {2002})}\BibitemShut {NoStop}%
\bibitem [{\citenamefont {Edwards}\ \emph {et~al.}(2001)\citenamefont
  {Edwards}, \citenamefont {Ibarra},\ and\ \citenamefont
  {Palsson}}]{edwards2001silico}%
  \BibitemOpen
  \bibfield  {author} {\bibinfo {author} {\bibfnamefont {J.~S.}\ \bibnamefont
  {Edwards}}, \bibinfo {author} {\bibfnamefont {R.~U.}\ \bibnamefont {Ibarra}},
  \ and\ \bibinfo {author} {\bibfnamefont {B.~O.}\ \bibnamefont {Palsson}},\
  }\href@noop {} {\bibfield  {journal} {\bibinfo  {journal} {Nature
  biotechnology}\ }\textbf {\bibinfo {volume} {19}},\ \bibinfo {pages} {125}
  (\bibinfo {year} {2001})}\BibitemShut {NoStop}%
\bibitem [{\citenamefont {Majewski}\ and\ \citenamefont
  {Domach}(1990)}]{majewski1990simple}%
  \BibitemOpen
  \bibfield  {author} {\bibinfo {author} {\bibfnamefont {R.}~\bibnamefont
  {Majewski}}\ and\ \bibinfo {author} {\bibfnamefont {M.}~\bibnamefont
  {Domach}},\ }\href@noop {} {\bibfield  {journal} {\bibinfo  {journal}
  {Biotechnology and bioengineering}\ }\textbf {\bibinfo {volume} {35}},\
  \bibinfo {pages} {732} (\bibinfo {year} {1990})}\BibitemShut {NoStop}%
\bibitem [{\citenamefont {Varma}\ and\ \citenamefont
  {Palsson}(1994)}]{Varma:1994gd}%
  \BibitemOpen
  \bibfield  {author} {\bibinfo {author} {\bibfnamefont {A.}~\bibnamefont
  {Varma}}\ and\ \bibinfo {author} {\bibfnamefont {B.}~\bibnamefont
  {Palsson}},\ }\href@noop {} {\bibfield  {journal} {\bibinfo  {journal}
  {Applied Environmental Microbiology}\ }\textbf {\bibinfo {volume} {60}},\
  \bibinfo {pages} {3724} (\bibinfo {year} {1994})}\BibitemShut {NoStop}%
\bibitem [{\citenamefont {Edwards}\ and\ \citenamefont
  {Palsson}(2000)}]{Edwards:2000zt}%
  \BibitemOpen
  \bibfield  {author} {\bibinfo {author} {\bibfnamefont {J.}~\bibnamefont
  {Edwards}}\ and\ \bibinfo {author} {\bibfnamefont {B.}~\bibnamefont
  {Palsson}},\ }\href {\doibase 10.1186/1471-2105-1-1} {\bibfield  {journal}
  {\bibinfo  {journal} {BMC Bioinformatics}\ }\textbf {\bibinfo {volume} {1}},\
  \bibinfo {pages} {1} (\bibinfo {year} {2000})}\BibitemShut {NoStop}%
\bibitem [{\citenamefont {Vazquez}\ \emph {et~al.}(2010)\citenamefont
  {Vazquez}, \citenamefont {Liu}, \citenamefont {Zhou},\ and\ \citenamefont
  {Oltvai}}]{vazquez2010catabolic}%
  \BibitemOpen
  \bibfield  {author} {\bibinfo {author} {\bibfnamefont {A.}~\bibnamefont
  {Vazquez}}, \bibinfo {author} {\bibfnamefont {J.}~\bibnamefont {Liu}},
  \bibinfo {author} {\bibfnamefont {Y.}~\bibnamefont {Zhou}}, \ and\ \bibinfo
  {author} {\bibfnamefont {Z.~N.}\ \bibnamefont {Oltvai}},\ }\href@noop {}
  {\bibfield  {journal} {\bibinfo  {journal} {BMC systems biology}\ }\textbf
  {\bibinfo {volume} {4}},\ \bibinfo {pages} {58} (\bibinfo {year}
  {2010})}\BibitemShut {NoStop}%
\bibitem [{\citenamefont {Wang}\ \emph {et~al.}(2010)\citenamefont {Wang},
  \citenamefont {Robert}, \citenamefont {Pelletier}, \citenamefont {Dang},
  \citenamefont {Taddei}, \citenamefont {Wright},\ and\ \citenamefont
  {Jun}}]{wang2010robust}%
  \BibitemOpen
  \bibfield  {author} {\bibinfo {author} {\bibfnamefont {P.}~\bibnamefont
  {Wang}}, \bibinfo {author} {\bibfnamefont {L.}~\bibnamefont {Robert}},
  \bibinfo {author} {\bibfnamefont {J.}~\bibnamefont {Pelletier}}, \bibinfo
  {author} {\bibfnamefont {W.~L.}\ \bibnamefont {Dang}}, \bibinfo {author}
  {\bibfnamefont {F.}~\bibnamefont {Taddei}}, \bibinfo {author} {\bibfnamefont
  {A.}~\bibnamefont {Wright}}, \ and\ \bibinfo {author} {\bibfnamefont
  {S.}~\bibnamefont {Jun}},\ }\href@noop {} {\bibfield  {journal} {\bibinfo
  {journal} {Current biology}\ }\textbf {\bibinfo {volume} {20}},\ \bibinfo
  {pages} {1099} (\bibinfo {year} {2010})}\BibitemShut {NoStop}%
\bibitem [{\citenamefont {Iyer-Biswas}\ \emph
  {et~al.}(2014{\natexlab{a}})\citenamefont {Iyer-Biswas}, \citenamefont
  {Crooks}, \citenamefont {Scherer},\ and\ \citenamefont
  {Dinner}}]{iyer2014universality}%
  \BibitemOpen
  \bibfield  {author} {\bibinfo {author} {\bibfnamefont {S.}~\bibnamefont
  {Iyer-Biswas}}, \bibinfo {author} {\bibfnamefont {G.~E.}\ \bibnamefont
  {Crooks}}, \bibinfo {author} {\bibfnamefont {N.~F.}\ \bibnamefont {Scherer}},
  \ and\ \bibinfo {author} {\bibfnamefont {A.~R.}\ \bibnamefont {Dinner}},\
  }\href@noop {} {\bibfield  {journal} {\bibinfo  {journal} {Physical review
  letters}\ }\textbf {\bibinfo {volume} {113}},\ \bibinfo {pages} {028101}
  (\bibinfo {year} {2014}{\natexlab{a}})}\BibitemShut {NoStop}%
\bibitem [{\citenamefont {Iyer-Biswas}\ \emph
  {et~al.}(2014{\natexlab{b}})\citenamefont {Iyer-Biswas}, \citenamefont
  {Wright}, \citenamefont {Henry}, \citenamefont {Lo}, \citenamefont {Burov},
  \citenamefont {Lin}, \citenamefont {Crooks}, \citenamefont {Crosson},
  \citenamefont {Dinner},\ and\ \citenamefont {Scherer}}]{iyer2014scaling}%
  \BibitemOpen
  \bibfield  {author} {\bibinfo {author} {\bibfnamefont {S.}~\bibnamefont
  {Iyer-Biswas}}, \bibinfo {author} {\bibfnamefont {C.~S.}\ \bibnamefont
  {Wright}}, \bibinfo {author} {\bibfnamefont {J.~T.}\ \bibnamefont {Henry}},
  \bibinfo {author} {\bibfnamefont {K.}~\bibnamefont {Lo}}, \bibinfo {author}
  {\bibfnamefont {S.}~\bibnamefont {Burov}}, \bibinfo {author} {\bibfnamefont
  {Y.}~\bibnamefont {Lin}}, \bibinfo {author} {\bibfnamefont {G.~E.}\
  \bibnamefont {Crooks}}, \bibinfo {author} {\bibfnamefont {S.}~\bibnamefont
  {Crosson}}, \bibinfo {author} {\bibfnamefont {A.~R.}\ \bibnamefont {Dinner}},
  \ and\ \bibinfo {author} {\bibfnamefont {N.~F.}\ \bibnamefont {Scherer}},\
  }\href@noop {} {\bibfield  {journal} {\bibinfo  {journal} {Proceedings of the
  National Academy of Sciences}\ }\textbf {\bibinfo {volume} {111}},\ \bibinfo
  {pages} {15912} (\bibinfo {year} {2014}{\natexlab{b}})}\BibitemShut {NoStop}%
\bibitem [{\citenamefont {Kennard}\ \emph {et~al.}(2016)\citenamefont
  {Kennard}, \citenamefont {Osella}, \citenamefont {Javer}, \citenamefont
  {Grilli}, \citenamefont {Nghe}, \citenamefont {Tans}, \citenamefont
  {Cicuta},\ and\ \citenamefont {Lagomarsino}}]{kennard2016individuality}%
  \BibitemOpen
  \bibfield  {author} {\bibinfo {author} {\bibfnamefont {A.~S.}\ \bibnamefont
  {Kennard}}, \bibinfo {author} {\bibfnamefont {M.}~\bibnamefont {Osella}},
  \bibinfo {author} {\bibfnamefont {A.}~\bibnamefont {Javer}}, \bibinfo
  {author} {\bibfnamefont {J.}~\bibnamefont {Grilli}}, \bibinfo {author}
  {\bibfnamefont {P.}~\bibnamefont {Nghe}}, \bibinfo {author} {\bibfnamefont
  {S.~J.}\ \bibnamefont {Tans}}, \bibinfo {author} {\bibfnamefont
  {P.}~\bibnamefont {Cicuta}}, \ and\ \bibinfo {author} {\bibfnamefont {M.~C.}\
  \bibnamefont {Lagomarsino}},\ }\href@noop {} {\bibfield  {journal} {\bibinfo
  {journal} {Physical Review E}\ }\textbf {\bibinfo {volume} {93}},\ \bibinfo
  {pages} {012408} (\bibinfo {year} {2016})}\BibitemShut {NoStop}%
\bibitem [{\citenamefont {Taheri-Araghi}\ \emph {et~al.}(2015)\citenamefont
  {Taheri-Araghi}, \citenamefont {Bradde}, \citenamefont {Sauls}, \citenamefont
  {Hill}, \citenamefont {Levin}, \citenamefont {Paulsson}, \citenamefont
  {Vergassola},\ and\ \citenamefont {Jun}}]{taheri2015cell}%
  \BibitemOpen
  \bibfield  {author} {\bibinfo {author} {\bibfnamefont {S.}~\bibnamefont
  {Taheri-Araghi}}, \bibinfo {author} {\bibfnamefont {S.}~\bibnamefont
  {Bradde}}, \bibinfo {author} {\bibfnamefont {J.~T.}\ \bibnamefont {Sauls}},
  \bibinfo {author} {\bibfnamefont {N.~S.}\ \bibnamefont {Hill}}, \bibinfo
  {author} {\bibfnamefont {P.~A.}\ \bibnamefont {Levin}}, \bibinfo {author}
  {\bibfnamefont {J.}~\bibnamefont {Paulsson}}, \bibinfo {author}
  {\bibfnamefont {M.}~\bibnamefont {Vergassola}}, \ and\ \bibinfo {author}
  {\bibfnamefont {S.}~\bibnamefont {Jun}},\ }\href@noop {} {\bibfield
  {journal} {\bibinfo  {journal} {Current Biology}\ }\textbf {\bibinfo {volume}
  {25}},\ \bibinfo {pages} {385} (\bibinfo {year} {2015})}\BibitemShut
  {NoStop}%
\bibitem [{\citenamefont {Kiviet}\ \emph {et~al.}(2014)\citenamefont {Kiviet},
  \citenamefont {Nghe}, \citenamefont {Walker}, \citenamefont {Boulineau},
  \citenamefont {Sunderlikova},\ and\ \citenamefont
  {Tans}}]{kiviet2014stochasticity}%
  \BibitemOpen
  \bibfield  {author} {\bibinfo {author} {\bibfnamefont {D.~J.}\ \bibnamefont
  {Kiviet}}, \bibinfo {author} {\bibfnamefont {P.}~\bibnamefont {Nghe}},
  \bibinfo {author} {\bibfnamefont {N.}~\bibnamefont {Walker}}, \bibinfo
  {author} {\bibfnamefont {S.}~\bibnamefont {Boulineau}}, \bibinfo {author}
  {\bibfnamefont {V.}~\bibnamefont {Sunderlikova}}, \ and\ \bibinfo {author}
  {\bibfnamefont {S.~J.}\ \bibnamefont {Tans}},\ }\href@noop {} {\bibfield
  {journal} {\bibinfo  {journal} {Nature}\ } (\bibinfo {year}
  {2014})}\BibitemShut {NoStop}%
\bibitem [{\citenamefont {Shahrezaei}\ and\ \citenamefont
  {Marguerat}(2015)}]{shahrezaei2015connecting}%
  \BibitemOpen
  \bibfield  {author} {\bibinfo {author} {\bibfnamefont {V.}~\bibnamefont
  {Shahrezaei}}\ and\ \bibinfo {author} {\bibfnamefont {S.}~\bibnamefont
  {Marguerat}},\ }\href@noop {} {\bibfield  {journal} {\bibinfo  {journal}
  {Current opinion in microbiology}\ }\textbf {\bibinfo {volume} {25}},\
  \bibinfo {pages} {127} (\bibinfo {year} {2015})}\BibitemShut {NoStop}%
\bibitem [{\citenamefont {Keren}\ \emph {et~al.}(2015)\citenamefont {Keren},
  \citenamefont {Van~Dijk}, \citenamefont {Weingarten-Gabbay}, \citenamefont
  {Davidi}, \citenamefont {Jona}, \citenamefont {Weinberger}, \citenamefont
  {Milo},\ and\ \citenamefont {Segal}}]{keren2015noise}%
  \BibitemOpen
  \bibfield  {author} {\bibinfo {author} {\bibfnamefont {L.}~\bibnamefont
  {Keren}}, \bibinfo {author} {\bibfnamefont {D.}~\bibnamefont {Van~Dijk}},
  \bibinfo {author} {\bibfnamefont {S.}~\bibnamefont {Weingarten-Gabbay}},
  \bibinfo {author} {\bibfnamefont {D.}~\bibnamefont {Davidi}}, \bibinfo
  {author} {\bibfnamefont {G.}~\bibnamefont {Jona}}, \bibinfo {author}
  {\bibfnamefont {A.}~\bibnamefont {Weinberger}}, \bibinfo {author}
  {\bibfnamefont {R.}~\bibnamefont {Milo}}, \ and\ \bibinfo {author}
  {\bibfnamefont {E.}~\bibnamefont {Segal}},\ }\href@noop {} {\bibfield
  {journal} {\bibinfo  {journal} {Genome research}\ ,\ \bibinfo {pages} {gr}}
  (\bibinfo {year} {2015})}\BibitemShut {NoStop}%
\bibitem [{\citenamefont {Cerulus}\ \emph {et~al.}(2016)\citenamefont
  {Cerulus}, \citenamefont {New}, \citenamefont {Pougach},\ and\ \citenamefont
  {Verstrepen}}]{cerulus2016noise}%
  \BibitemOpen
  \bibfield  {author} {\bibinfo {author} {\bibfnamefont {B.}~\bibnamefont
  {Cerulus}}, \bibinfo {author} {\bibfnamefont {A.~M.}\ \bibnamefont {New}},
  \bibinfo {author} {\bibfnamefont {K.}~\bibnamefont {Pougach}}, \ and\
  \bibinfo {author} {\bibfnamefont {K.~J.}\ \bibnamefont {Verstrepen}},\
  }\href@noop {} {\bibfield  {journal} {\bibinfo  {journal} {Current Biology}\
  }\textbf {\bibinfo {volume} {26}},\ \bibinfo {pages} {1138} (\bibinfo {year}
  {2016})}\BibitemShut {NoStop}%
\bibitem [{\citenamefont {Jaynes}(1957)}]{jaynes1957information}%
  \BibitemOpen
  \bibfield  {author} {\bibinfo {author} {\bibfnamefont {E.~T.}\ \bibnamefont
  {Jaynes}},\ }\href@noop {} {\bibfield  {journal} {\bibinfo  {journal}
  {Physical review}\ }\textbf {\bibinfo {volume} {106}},\ \bibinfo {pages}
  {620} (\bibinfo {year} {1957})}\BibitemShut {NoStop}%
\bibitem [{\citenamefont {Schneidman}\ \emph {et~al.}(2006)\citenamefont
  {Schneidman}, \citenamefont {Berry}, \citenamefont {Segev},\ and\
  \citenamefont {Bialek}}]{schneidman2006weak}%
  \BibitemOpen
  \bibfield  {author} {\bibinfo {author} {\bibfnamefont {E.}~\bibnamefont
  {Schneidman}}, \bibinfo {author} {\bibfnamefont {M.~J.}\ \bibnamefont
  {Berry}}, \bibinfo {author} {\bibfnamefont {R.}~\bibnamefont {Segev}}, \ and\
  \bibinfo {author} {\bibfnamefont {W.}~\bibnamefont {Bialek}},\ }\href@noop {}
  {\bibfield  {journal} {\bibinfo  {journal} {Nature}\ }\textbf {\bibinfo
  {volume} {440}},\ \bibinfo {pages} {1007} (\bibinfo {year}
  {2006})}\BibitemShut {NoStop}%
\bibitem [{\citenamefont {Tka{\v{c}}ik}\ \emph {et~al.}(2014)\citenamefont
  {Tka{\v{c}}ik}, \citenamefont {Marre}, \citenamefont {Amodei}, \citenamefont
  {Schneidman}, \citenamefont {Bialek},\ and\ \citenamefont
  {Berry~II}}]{tkacikpcb}%
  \BibitemOpen
  \bibfield  {author} {\bibinfo {author} {\bibfnamefont {G.}~\bibnamefont
  {Tka{\v{c}}ik}}, \bibinfo {author} {\bibfnamefont {O.}~\bibnamefont {Marre}},
  \bibinfo {author} {\bibfnamefont {D.}~\bibnamefont {Amodei}}, \bibinfo
  {author} {\bibfnamefont {E.}~\bibnamefont {Schneidman}}, \bibinfo {author}
  {\bibfnamefont {W.}~\bibnamefont {Bialek}}, \ and\ \bibinfo {author}
  {\bibfnamefont {M.~J.}\ \bibnamefont {Berry~II}},\ }\href@noop {} {\bibfield
  {journal} {\bibinfo  {journal} {PLoS Comput Biol}\ }\textbf {\bibinfo
  {volume} {10}},\ \bibinfo {pages} {e1003408} (\bibinfo {year}
  {2014})}\BibitemShut {NoStop}%
\bibitem [{\citenamefont {Lezon}\ \emph {et~al.}(2006)\citenamefont {Lezon},
  \citenamefont {Banavar}, \citenamefont {Cieplak}, \citenamefont {Maritan},\
  and\ \citenamefont {Fedoroff}}]{lezon2006using}%
  \BibitemOpen
  \bibfield  {author} {\bibinfo {author} {\bibfnamefont {T.~R.}\ \bibnamefont
  {Lezon}}, \bibinfo {author} {\bibfnamefont {J.~R.}\ \bibnamefont {Banavar}},
  \bibinfo {author} {\bibfnamefont {M.}~\bibnamefont {Cieplak}}, \bibinfo
  {author} {\bibfnamefont {A.}~\bibnamefont {Maritan}}, \ and\ \bibinfo
  {author} {\bibfnamefont {N.~V.}\ \bibnamefont {Fedoroff}},\ }\href@noop {}
  {\bibfield  {journal} {\bibinfo  {journal} {Proceedings of the National
  Academy of Sciences}\ }\textbf {\bibinfo {volume} {103}},\ \bibinfo {pages}
  {19033} (\bibinfo {year} {2006})}\BibitemShut {NoStop}%
\bibitem [{\citenamefont {Mora}\ \emph {et~al.}(2010)\citenamefont {Mora},
  \citenamefont {Walczak}, \citenamefont {Bialek},\ and\ \citenamefont
  {Callan}}]{mora2010maximum}%
  \BibitemOpen
  \bibfield  {author} {\bibinfo {author} {\bibfnamefont {T.}~\bibnamefont
  {Mora}}, \bibinfo {author} {\bibfnamefont {A.~M.}\ \bibnamefont {Walczak}},
  \bibinfo {author} {\bibfnamefont {W.}~\bibnamefont {Bialek}}, \ and\ \bibinfo
  {author} {\bibfnamefont {C.~G.}\ \bibnamefont {Callan}},\ }\href@noop {}
  {\bibfield  {journal} {\bibinfo  {journal} {Proceedings of the National
  Academy of Sciences}\ }\textbf {\bibinfo {volume} {107}},\ \bibinfo {pages}
  {5405} (\bibinfo {year} {2010})}\BibitemShut {NoStop}%
\bibitem [{\citenamefont {Bialek}\ \emph {et~al.}(2012)\citenamefont {Bialek},
  \citenamefont {Cavagna}, \citenamefont {Giardina}, \citenamefont {Mora},
  \citenamefont {Silvestri}, \citenamefont {Viale},\ and\ \citenamefont
  {Walczak}}]{bialek2012statistical}%
  \BibitemOpen
  \bibfield  {author} {\bibinfo {author} {\bibfnamefont {W.}~\bibnamefont
  {Bialek}}, \bibinfo {author} {\bibfnamefont {A.}~\bibnamefont {Cavagna}},
  \bibinfo {author} {\bibfnamefont {I.}~\bibnamefont {Giardina}}, \bibinfo
  {author} {\bibfnamefont {T.}~\bibnamefont {Mora}}, \bibinfo {author}
  {\bibfnamefont {E.}~\bibnamefont {Silvestri}}, \bibinfo {author}
  {\bibfnamefont {M.}~\bibnamefont {Viale}}, \ and\ \bibinfo {author}
  {\bibfnamefont {A.~M.}\ \bibnamefont {Walczak}},\ }\href@noop {} {\bibfield
  {journal} {\bibinfo  {journal} {Proceedings of the National Academy of
  Sciences}\ }\textbf {\bibinfo {volume} {109}},\ \bibinfo {pages} {4786}
  (\bibinfo {year} {2012})}\BibitemShut {NoStop}%
\bibitem [{\citenamefont {De~Martino}\ \emph {et~al.}(2016)\citenamefont
  {De~Martino}, \citenamefont {Capuani},\ and\ \citenamefont
  {De~Martino}}]{de2016growth}%
  \BibitemOpen
  \bibfield  {author} {\bibinfo {author} {\bibfnamefont {D.}~\bibnamefont
  {De~Martino}}, \bibinfo {author} {\bibfnamefont {F.}~\bibnamefont {Capuani}},
  \ and\ \bibinfo {author} {\bibfnamefont {A.}~\bibnamefont {De~Martino}},\
  }\href@noop {} {\bibfield  {journal} {\bibinfo  {journal} {Physical biology}\
  }\textbf {\bibinfo {volume} {13}},\ \bibinfo {pages} {036005} (\bibinfo
  {year} {2016})}\BibitemShut {NoStop}%
\bibitem [{\citenamefont {Zhang}\ \emph
  {et~al.}(2014{\natexlab{a}})\citenamefont {Zhang}, \citenamefont {Shen},
  \citenamefont {Rui}, \citenamefont {Zhou}, \citenamefont {Zhou},
  \citenamefont {Shang}, \citenamefont {Xin}, \citenamefont {Liu},
  \citenamefont {Li}, \citenamefont {Jiang} \emph {et~al.}}]{cecafdb}%
  \BibitemOpen
  \bibfield  {author} {\bibinfo {author} {\bibfnamefont {Z.}~\bibnamefont
  {Zhang}}, \bibinfo {author} {\bibfnamefont {T.}~\bibnamefont {Shen}},
  \bibinfo {author} {\bibfnamefont {B.}~\bibnamefont {Rui}}, \bibinfo {author}
  {\bibfnamefont {W.}~\bibnamefont {Zhou}}, \bibinfo {author} {\bibfnamefont
  {X.}~\bibnamefont {Zhou}}, \bibinfo {author} {\bibfnamefont {C.}~\bibnamefont
  {Shang}}, \bibinfo {author} {\bibfnamefont {C.}~\bibnamefont {Xin}}, \bibinfo
  {author} {\bibfnamefont {X.}~\bibnamefont {Liu}}, \bibinfo {author}
  {\bibfnamefont {G.}~\bibnamefont {Li}}, \bibinfo {author} {\bibfnamefont
  {J.}~\bibnamefont {Jiang}},  \emph {et~al.},\ }\href@noop {} {\bibfield
  {journal} {\bibinfo  {journal} {Nucleic acids research}\ ,\ \bibinfo {pages}
  {gku1137}} (\bibinfo {year} {2014}{\natexlab{a}})}\BibitemShut {NoStop}%
\bibitem [{\citenamefont {Ishii}\ \emph {et~al.}(2007)\citenamefont {Ishii},
  \citenamefont {Nakahigashi}, \citenamefont {Baba}, \citenamefont {Robert},
  \citenamefont {Soga}, \citenamefont {Kanai}, \citenamefont {Hirasawa},
  \citenamefont {Naba}, \citenamefont {Hirai}, \citenamefont {Hoque} \emph
  {et~al.}}]{ishii2007multiple}%
  \BibitemOpen
  \bibfield  {author} {\bibinfo {author} {\bibfnamefont {N.}~\bibnamefont
  {Ishii}}, \bibinfo {author} {\bibfnamefont {K.}~\bibnamefont {Nakahigashi}},
  \bibinfo {author} {\bibfnamefont {T.}~\bibnamefont {Baba}}, \bibinfo {author}
  {\bibfnamefont {M.}~\bibnamefont {Robert}}, \bibinfo {author} {\bibfnamefont
  {T.}~\bibnamefont {Soga}}, \bibinfo {author} {\bibfnamefont {A.}~\bibnamefont
  {Kanai}}, \bibinfo {author} {\bibfnamefont {T.}~\bibnamefont {Hirasawa}},
  \bibinfo {author} {\bibfnamefont {M.}~\bibnamefont {Naba}}, \bibinfo {author}
  {\bibfnamefont {K.}~\bibnamefont {Hirai}}, \bibinfo {author} {\bibfnamefont
  {A.}~\bibnamefont {Hoque}},  \emph {et~al.},\ }\href@noop {} {\bibfield
  {journal} {\bibinfo  {journal} {Science}\ }\textbf {\bibinfo {volume}
  {316}},\ \bibinfo {pages} {593} (\bibinfo {year} {2007})}\BibitemShut
  {NoStop}%
\bibitem [{\citenamefont {Orth}\ \emph {et~al.}(2011)\citenamefont {Orth},
  \citenamefont {Conrad}, \citenamefont {Na}, \citenamefont {Lerman},
  \citenamefont {Nam}, \citenamefont {Feist},\ and\ \citenamefont
  {Palsson}}]{orth2011comprehensive}%
  \BibitemOpen
  \bibfield  {author} {\bibinfo {author} {\bibfnamefont {J.~D.}\ \bibnamefont
  {Orth}}, \bibinfo {author} {\bibfnamefont {T.~M.}\ \bibnamefont {Conrad}},
  \bibinfo {author} {\bibfnamefont {J.}~\bibnamefont {Na}}, \bibinfo {author}
  {\bibfnamefont {J.~A.}\ \bibnamefont {Lerman}}, \bibinfo {author}
  {\bibfnamefont {H.}~\bibnamefont {Nam}}, \bibinfo {author} {\bibfnamefont
  {A.~M.}\ \bibnamefont {Feist}}, \ and\ \bibinfo {author} {\bibfnamefont
  {B.~{\O}.}\ \bibnamefont {Palsson}},\ }\href@noop {} {\bibfield  {journal}
  {\bibinfo  {journal} {Molecular systems biology}\ }\textbf {\bibinfo {volume}
  {7}},\ \bibinfo {pages} {535} (\bibinfo {year} {2011})}\BibitemShut {NoStop}%
\bibitem [{\citenamefont {De~Martino}\ \emph {et~al.}(2015)\citenamefont
  {De~Martino}, \citenamefont {Mori},\ and\ \citenamefont
  {Parisi}}]{de2015uniform}%
  \BibitemOpen
  \bibfield  {author} {\bibinfo {author} {\bibfnamefont {D.}~\bibnamefont
  {De~Martino}}, \bibinfo {author} {\bibfnamefont {M.}~\bibnamefont {Mori}}, \
  and\ \bibinfo {author} {\bibfnamefont {V.}~\bibnamefont {Parisi}},\
  }\href@noop {} {\bibfield  {journal} {\bibinfo  {journal} {PloS one}\
  }\textbf {\bibinfo {volume} {10}},\ \bibinfo {pages} {e0122670} (\bibinfo
  {year} {2015})}\BibitemShut {NoStop}%
\bibitem [{\citenamefont {Slonim}\ \emph {et~al.}(2005)\citenamefont {Slonim},
  \citenamefont {Atwal}, \citenamefont {Tka{\v{c}}ik},\ and\ \citenamefont
  {Bialek}}]{iclust}%
  \BibitemOpen
  \bibfield  {author} {\bibinfo {author} {\bibfnamefont {N.}~\bibnamefont
  {Slonim}}, \bibinfo {author} {\bibfnamefont {G.~S.}\ \bibnamefont {Atwal}},
  \bibinfo {author} {\bibfnamefont {G.}~\bibnamefont {Tka{\v{c}}ik}}, \ and\
  \bibinfo {author} {\bibfnamefont {W.}~\bibnamefont {Bialek}},\ }\href@noop {}
  {\bibfield  {journal} {\bibinfo  {journal} {Proceedings of the National
  Academy of Sciences of the United States of America}\ }\textbf {\bibinfo
  {volume} {102}},\ \bibinfo {pages} {18297} (\bibinfo {year}
  {2005})}\BibitemShut {NoStop}%
\bibitem [{\citenamefont {Tka{\v{c}}ik}\ and\ \citenamefont
  {Walczak}(2011)}]{gtawreview}%
  \BibitemOpen
  \bibfield  {author} {\bibinfo {author} {\bibfnamefont {G.}~\bibnamefont
  {Tka{\v{c}}ik}}\ and\ \bibinfo {author} {\bibfnamefont {A.~M.}\ \bibnamefont
  {Walczak}},\ }\href@noop {} {\bibfield  {journal} {\bibinfo  {journal}
  {Journal of Physics: Condensed Matter}\ }\textbf {\bibinfo {volume} {23}},\
  \bibinfo {pages} {153102} (\bibinfo {year} {2011})}\BibitemShut {NoStop}%
\bibitem [{\citenamefont {Tka{\v{c}}ik}\ and\ \citenamefont
  {Bialek}(2016)}]{annurevgt}%
  \BibitemOpen
  \bibfield  {author} {\bibinfo {author} {\bibfnamefont {G.}~\bibnamefont
  {Tka{\v{c}}ik}}\ and\ \bibinfo {author} {\bibfnamefont {W.}~\bibnamefont
  {Bialek}},\ }\href@noop {} {\bibfield  {journal} {\bibinfo  {journal} {Annual
  Review of Condensed Matter Physics}\ }\textbf {\bibinfo {volume} {7}},\
  \bibinfo {pages} {89} (\bibinfo {year} {2016})}\BibitemShut {NoStop}%
\bibitem [{\citenamefont {Bergmiller}\ \emph {et~al.}(2017)\citenamefont
  {Bergmiller}, \citenamefont {Andersson}, \citenamefont {Tomasek},
  \citenamefont {Balleza}, \citenamefont {Kiviet}, \citenamefont {Hauschild},
  \citenamefont {Tka\v{c}ik},\ and\ \citenamefont {Guet}}]{tobianna}%
  \BibitemOpen
  \bibfield  {author} {\bibinfo {author} {\bibfnamefont {T.}~\bibnamefont
  {Bergmiller}}, \bibinfo {author} {\bibfnamefont {A.}~\bibnamefont
  {Andersson}}, \bibinfo {author} {\bibfnamefont {K.}~\bibnamefont {Tomasek}},
  \bibinfo {author} {\bibfnamefont {E.}~\bibnamefont {Balleza}}, \bibinfo
  {author} {\bibfnamefont {D.}~\bibnamefont {Kiviet}}, \bibinfo {author}
  {\bibfnamefont {R.}~\bibnamefont {Hauschild}}, \bibinfo {author}
  {\bibfnamefont {G.}~\bibnamefont {Tka\v{c}ik}}, \ and\ \bibinfo {author}
  {\bibfnamefont {C.}~\bibnamefont {Guet}},\ }\href@noop {} {\bibfield
  {journal} {\bibinfo  {journal} {to appear}\ } (\bibinfo {year}
  {2017})}\BibitemShut {NoStop}%
\bibitem [{\citenamefont {De~Martino}\ and\ \citenamefont
  {Masoero}(2016)}]{demaso}%
  \BibitemOpen
  \bibfield  {author} {\bibinfo {author} {\bibfnamefont {D.}~\bibnamefont
  {De~Martino}}\ and\ \bibinfo {author} {\bibfnamefont {D.}~\bibnamefont
  {Masoero}},\ }\href@noop {} {\bibfield  {journal} {\bibinfo  {journal}
  {Journal of Statistical Mechanics: Theory and Experiment}\ }\textbf {\bibinfo
  {volume} {2016}},\ \bibinfo {pages} {123502} (\bibinfo {year}
  {2016})}\BibitemShut {NoStop}%
\bibitem [{\citenamefont {Mori}\ \emph {et~al.}(2016)\citenamefont {Mori},
  \citenamefont {Hwa}, \citenamefont {Martin}, \citenamefont {De~Martino},\
  and\ \citenamefont {Marinari}}]{mori2016constrained}%
  \BibitemOpen
  \bibfield  {author} {\bibinfo {author} {\bibfnamefont {M.}~\bibnamefont
  {Mori}}, \bibinfo {author} {\bibfnamefont {T.}~\bibnamefont {Hwa}}, \bibinfo
  {author} {\bibfnamefont {O.~C.}\ \bibnamefont {Martin}}, \bibinfo {author}
  {\bibfnamefont {A.}~\bibnamefont {De~Martino}}, \ and\ \bibinfo {author}
  {\bibfnamefont {E.}~\bibnamefont {Marinari}},\ }\href@noop {} {\bibfield
  {journal} {\bibinfo  {journal} {PLoS Comput Biol}\ }\textbf {\bibinfo
  {volume} {12}},\ \bibinfo {pages} {e1004913} (\bibinfo {year}
  {2016})}\BibitemShut {NoStop}%
\bibitem [{\citenamefont {Dyer}\ and\ \citenamefont
  {Frieze}(1988)}]{dyer1988complexity}%
  \BibitemOpen
  \bibfield  {author} {\bibinfo {author} {\bibfnamefont {M.~E.}\ \bibnamefont
  {Dyer}}\ and\ \bibinfo {author} {\bibfnamefont {A.~M.}\ \bibnamefont
  {Frieze}},\ }\href@noop {} {\bibfield  {journal} {\bibinfo  {journal} {SIAM
  Journal on Computing}\ }\textbf {\bibinfo {volume} {17}},\ \bibinfo {pages}
  {967} (\bibinfo {year} {1988})}\BibitemShut {NoStop}%
\bibitem [{\citenamefont {Turcin}(1971)}]{Turcin:1971}%
  \BibitemOpen
  \bibfield  {author} {\bibinfo {author} {\bibfnamefont {V.}~\bibnamefont
  {Turcin}},\ }\href@noop {} {\bibfield  {journal} {\bibinfo  {journal} {Th
  Probab Appl}\ }\textbf {\bibinfo {volume} {16}},\ \bibinfo {pages} {720}
  (\bibinfo {year} {1971})}\BibitemShut {NoStop}%
\bibitem [{\citenamefont {Smith}(1996)}]{Smith:1996p4127}%
  \BibitemOpen
  \bibfield  {author} {\bibinfo {author} {\bibfnamefont {R.}~\bibnamefont
  {Smith}},\ }\href@noop {} {\bibfield  {journal} {\bibinfo  {journal} {Proc
  1996 Winter Simul Conf}\ ,\ \bibinfo {pages} {260}} (\bibinfo {year}
  {1996})}\BibitemShut {NoStop}%
\bibitem [{\citenamefont {Lov{\'a}sz}(1999)}]{Lovasz:1999p4121}%
  \BibitemOpen
  \bibfield  {author} {\bibinfo {author} {\bibfnamefont {L.}~\bibnamefont
  {Lov{\'a}sz}},\ }\href@noop {} {\bibfield  {journal} {\bibinfo  {journal}
  {Math Program}\ }\textbf {\bibinfo {volume} {86}},\ \bibinfo {pages} {443}
  (\bibinfo {year} {1999})}\BibitemShut {NoStop}%
\bibitem [{\citenamefont {Ball}(1997)}]{ball1997elementary}%
  \BibitemOpen
  \bibfield  {author} {\bibinfo {author} {\bibfnamefont {K.}~\bibnamefont
  {Ball}},\ }\href@noop {} {\bibfield  {journal} {\bibinfo  {journal} {Flavors
  of geometry}\ }\textbf {\bibinfo {volume} {31}},\ \bibinfo {pages} {1}
  (\bibinfo {year} {1997})}\BibitemShut {NoStop}%
\bibitem [{\citenamefont {Lov{\'a}sz}(1987)}]{lovasz1987algorithmic}%
  \BibitemOpen
  \bibfield  {author} {\bibinfo {author} {\bibfnamefont {L.}~\bibnamefont
  {Lov{\'a}sz}},\ }\href@noop {} {\emph {\bibinfo {title} {An algorithmic
  theory of numbers, graphs and convexity}}},\ Vol.~\bibinfo {volume} {50}\
  (\bibinfo  {publisher} {SIAM},\ \bibinfo {year} {1987})\BibitemShut {NoStop}%
\bibitem [{\citenamefont {Bland}\ \emph {et~al.}(1981)\citenamefont {Bland},
  \citenamefont {Goldfarb},\ and\ \citenamefont {Todd}}]{bland1981ellipsoid}%
  \BibitemOpen
  \bibfield  {author} {\bibinfo {author} {\bibfnamefont {R.~G.}\ \bibnamefont
  {Bland}}, \bibinfo {author} {\bibfnamefont {D.}~\bibnamefont {Goldfarb}}, \
  and\ \bibinfo {author} {\bibfnamefont {M.~J.}\ \bibnamefont {Todd}},\
  }\href@noop {} {\bibfield  {journal} {\bibinfo  {journal} {Operations
  research}\ }\textbf {\bibinfo {volume} {29}},\ \bibinfo {pages} {1039}
  (\bibinfo {year} {1981})}\BibitemShut {NoStop}%
\bibitem [{\citenamefont {Zhang}\ \emph
  {et~al.}(2014{\natexlab{b}})\citenamefont {Zhang}, \citenamefont {Shen},
  \citenamefont {Rui}, \citenamefont {Zhou}, \citenamefont {Zhou},
  \citenamefont {Shang}, \citenamefont {Xin}, \citenamefont {Liu},
  \citenamefont {Li}, \citenamefont {Jiang} \emph {et~al.}}]{zhang2014cecafdb}%
  \BibitemOpen
  \bibfield  {author} {\bibinfo {author} {\bibfnamefont {Z.}~\bibnamefont
  {Zhang}}, \bibinfo {author} {\bibfnamefont {T.}~\bibnamefont {Shen}},
  \bibinfo {author} {\bibfnamefont {B.}~\bibnamefont {Rui}}, \bibinfo {author}
  {\bibfnamefont {W.}~\bibnamefont {Zhou}}, \bibinfo {author} {\bibfnamefont
  {X.}~\bibnamefont {Zhou}}, \bibinfo {author} {\bibfnamefont {C.}~\bibnamefont
  {Shang}}, \bibinfo {author} {\bibfnamefont {C.}~\bibnamefont {Xin}}, \bibinfo
  {author} {\bibfnamefont {X.}~\bibnamefont {Liu}}, \bibinfo {author}
  {\bibfnamefont {G.}~\bibnamefont {Li}}, \bibinfo {author} {\bibfnamefont
  {J.}~\bibnamefont {Jiang}},  \emph {et~al.},\ }\href@noop {} {\bibfield
  {journal} {\bibinfo  {journal} {Nucleic acids research}\ ,\ \bibinfo {pages}
  {gku1137}} (\bibinfo {year} {2014}{\natexlab{b}})}\BibitemShut {NoStop}%
\bibitem [{\citenamefont {Farmer}\ and\ \citenamefont
  {Liao}(1997)}]{farmer1997reduction}%
  \BibitemOpen
  \bibfield  {author} {\bibinfo {author} {\bibfnamefont {W.~R.}\ \bibnamefont
  {Farmer}}\ and\ \bibinfo {author} {\bibfnamefont {J.~C.}\ \bibnamefont
  {Liao}},\ }\href@noop {} {\bibfield  {journal} {\bibinfo  {journal} {Applied
  and environmental microbiology}\ }\textbf {\bibinfo {volume} {63}},\ \bibinfo
  {pages} {3205} (\bibinfo {year} {1997})}\BibitemShut {NoStop}%
\bibitem [{\citenamefont {Siddiquee}\ \emph {et~al.}(2004)\citenamefont
  {Siddiquee}, \citenamefont {Arauzo-Bravo},\ and\ \citenamefont
  {Shimizu}}]{siddiquee2004metabolic}%
  \BibitemOpen
  \bibfield  {author} {\bibinfo {author} {\bibfnamefont {K.~A.~Z.}\
  \bibnamefont {Siddiquee}}, \bibinfo {author} {\bibfnamefont {M.}~\bibnamefont
  {Arauzo-Bravo}}, \ and\ \bibinfo {author} {\bibfnamefont {K.}~\bibnamefont
  {Shimizu}},\ }\href@noop {} {\bibfield  {journal} {\bibinfo  {journal}
  {Applied Microbiology and Biotechnology}\ }\textbf {\bibinfo {volume} {63}},\
  \bibinfo {pages} {407} (\bibinfo {year} {2004})}\BibitemShut {NoStop}%
\bibitem [{\citenamefont {Schaub}\ \emph {et~al.}(2008)\citenamefont {Schaub},
  \citenamefont {Mauch},\ and\ \citenamefont {Reuss}}]{schaub2008metabolic}%
  \BibitemOpen
  \bibfield  {author} {\bibinfo {author} {\bibfnamefont {J.}~\bibnamefont
  {Schaub}}, \bibinfo {author} {\bibfnamefont {K.}~\bibnamefont {Mauch}}, \
  and\ \bibinfo {author} {\bibfnamefont {M.}~\bibnamefont {Reuss}},\
  }\href@noop {} {\bibfield  {journal} {\bibinfo  {journal} {Biotechnology and
  bioengineering}\ }\textbf {\bibinfo {volume} {99}},\ \bibinfo {pages} {1170}
  (\bibinfo {year} {2008})}\BibitemShut {NoStop}%
\bibitem [{\citenamefont {Peng}\ \emph {et~al.}(2004)\citenamefont {Peng},
  \citenamefont {Arauzo-Bravo},\ and\ \citenamefont
  {Shimizu}}]{peng2004metabolic}%
  \BibitemOpen
  \bibfield  {author} {\bibinfo {author} {\bibfnamefont {L.}~\bibnamefont
  {Peng}}, \bibinfo {author} {\bibfnamefont {M.~J.}\ \bibnamefont
  {Arauzo-Bravo}}, \ and\ \bibinfo {author} {\bibfnamefont {K.}~\bibnamefont
  {Shimizu}},\ }\href@noop {} {\bibfield  {journal} {\bibinfo  {journal} {FEMS
  microbiology letters}\ }\textbf {\bibinfo {volume} {235}},\ \bibinfo {pages}
  {17} (\bibinfo {year} {2004})}\BibitemShut {NoStop}%
\bibitem [{\citenamefont {Emmerling}\ \emph {et~al.}(2002)\citenamefont
  {Emmerling}, \citenamefont {Dauner}, \citenamefont {Ponti}, \citenamefont
  {Fiaux}, \citenamefont {Hochuli}, \citenamefont {Szyperski}, \citenamefont
  {W{\"u}thrich}, \citenamefont {Bailey},\ and\ \citenamefont
  {Sauer}}]{emmerling2002metabolic}%
  \BibitemOpen
  \bibfield  {author} {\bibinfo {author} {\bibfnamefont {M.}~\bibnamefont
  {Emmerling}}, \bibinfo {author} {\bibfnamefont {M.}~\bibnamefont {Dauner}},
  \bibinfo {author} {\bibfnamefont {A.}~\bibnamefont {Ponti}}, \bibinfo
  {author} {\bibfnamefont {J.}~\bibnamefont {Fiaux}}, \bibinfo {author}
  {\bibfnamefont {M.}~\bibnamefont {Hochuli}}, \bibinfo {author} {\bibfnamefont
  {T.}~\bibnamefont {Szyperski}}, \bibinfo {author} {\bibfnamefont
  {K.}~\bibnamefont {W{\"u}thrich}}, \bibinfo {author} {\bibfnamefont
  {J.}~\bibnamefont {Bailey}}, \ and\ \bibinfo {author} {\bibfnamefont
  {U.}~\bibnamefont {Sauer}},\ }\href@noop {} {\bibfield  {journal} {\bibinfo
  {journal} {Journal of bacteriology}\ }\textbf {\bibinfo {volume} {184}},\
  \bibinfo {pages} {152} (\bibinfo {year} {2002})}\BibitemShut {NoStop}%
\bibitem [{\citenamefont {Hua}\ \emph {et~al.}(2003)\citenamefont {Hua},
  \citenamefont {Yang}, \citenamefont {Baba}, \citenamefont {Mori},\ and\
  \citenamefont {Shimizu}}]{hua2003responses}%
  \BibitemOpen
  \bibfield  {author} {\bibinfo {author} {\bibfnamefont {Q.}~\bibnamefont
  {Hua}}, \bibinfo {author} {\bibfnamefont {C.}~\bibnamefont {Yang}}, \bibinfo
  {author} {\bibfnamefont {T.}~\bibnamefont {Baba}}, \bibinfo {author}
  {\bibfnamefont {H.}~\bibnamefont {Mori}}, \ and\ \bibinfo {author}
  {\bibfnamefont {K.}~\bibnamefont {Shimizu}},\ }\href@noop {} {\bibfield
  {journal} {\bibinfo  {journal} {Journal of Bacteriology}\ }\textbf {\bibinfo
  {volume} {185}},\ \bibinfo {pages} {7053} (\bibinfo {year}
  {2003})}\BibitemShut {NoStop}%
\bibitem [{\citenamefont {Nanchen}\ \emph {et~al.}(2008)\citenamefont
  {Nanchen}, \citenamefont {Schicker}, \citenamefont {Revelles},\ and\
  \citenamefont {Sauer}}]{nanchen2008cyclic}%
  \BibitemOpen
  \bibfield  {author} {\bibinfo {author} {\bibfnamefont {A.}~\bibnamefont
  {Nanchen}}, \bibinfo {author} {\bibfnamefont {A.}~\bibnamefont {Schicker}},
  \bibinfo {author} {\bibfnamefont {O.}~\bibnamefont {Revelles}}, \ and\
  \bibinfo {author} {\bibfnamefont {U.}~\bibnamefont {Sauer}},\ }\href@noop {}
  {\bibfield  {journal} {\bibinfo  {journal} {Journal of bacteriology}\
  }\textbf {\bibinfo {volume} {190}},\ \bibinfo {pages} {2323} (\bibinfo {year}
  {2008})}\BibitemShut {NoStop}%
\bibitem [{\citenamefont {Nikel}\ \emph {et~al.}(2009)\citenamefont {Nikel},
  \citenamefont {Zhu}, \citenamefont {San}, \citenamefont {M{\'e}ndez},\ and\
  \citenamefont {Bennett}}]{nikel2009metabolic}%
  \BibitemOpen
  \bibfield  {author} {\bibinfo {author} {\bibfnamefont {P.~I.}\ \bibnamefont
  {Nikel}}, \bibinfo {author} {\bibfnamefont {J.}~\bibnamefont {Zhu}}, \bibinfo
  {author} {\bibfnamefont {K.-Y.}\ \bibnamefont {San}}, \bibinfo {author}
  {\bibfnamefont {B.~S.}\ \bibnamefont {M{\'e}ndez}}, \ and\ \bibinfo {author}
  {\bibfnamefont {G.~N.}\ \bibnamefont {Bennett}},\ }\href@noop {} {\bibfield
  {journal} {\bibinfo  {journal} {Journal of bacteriology}\ }\textbf {\bibinfo
  {volume} {191}},\ \bibinfo {pages} {5538} (\bibinfo {year}
  {2009})}\BibitemShut {NoStop}%
\bibitem [{\citenamefont {Li}\ \emph {et~al.}(2006)\citenamefont {Li},
  \citenamefont {Ho}, \citenamefont {Yao},\ and\ \citenamefont
  {Shimizu}}]{li2006effect}%
  \BibitemOpen
  \bibfield  {author} {\bibinfo {author} {\bibfnamefont {M.}~\bibnamefont
  {Li}}, \bibinfo {author} {\bibfnamefont {P.~Y.}\ \bibnamefont {Ho}}, \bibinfo
  {author} {\bibfnamefont {S.}~\bibnamefont {Yao}}, \ and\ \bibinfo {author}
  {\bibfnamefont {K.}~\bibnamefont {Shimizu}},\ }\href@noop {} {\bibfield
  {journal} {\bibinfo  {journal} {Journal of biotechnology}\ }\textbf {\bibinfo
  {volume} {122}},\ \bibinfo {pages} {254} (\bibinfo {year}
  {2006})}\BibitemShut {NoStop}%
\bibitem [{\citenamefont {Toya}\ \emph {et~al.}(2007)\citenamefont {Toya},
  \citenamefont {Ishii}, \citenamefont {Hirasawa}, \citenamefont {Naba},
  \citenamefont {Hirai}, \citenamefont {Sugawara}, \citenamefont {Igarashi},
  \citenamefont {Shimizu}, \citenamefont {Tomita},\ and\ \citenamefont
  {Soga}}]{toya2007direct}%
  \BibitemOpen
  \bibfield  {author} {\bibinfo {author} {\bibfnamefont {Y.}~\bibnamefont
  {Toya}}, \bibinfo {author} {\bibfnamefont {N.}~\bibnamefont {Ishii}},
  \bibinfo {author} {\bibfnamefont {T.}~\bibnamefont {Hirasawa}}, \bibinfo
  {author} {\bibfnamefont {M.}~\bibnamefont {Naba}}, \bibinfo {author}
  {\bibfnamefont {K.}~\bibnamefont {Hirai}}, \bibinfo {author} {\bibfnamefont
  {K.}~\bibnamefont {Sugawara}}, \bibinfo {author} {\bibfnamefont
  {S.}~\bibnamefont {Igarashi}}, \bibinfo {author} {\bibfnamefont
  {K.}~\bibnamefont {Shimizu}}, \bibinfo {author} {\bibfnamefont
  {M.}~\bibnamefont {Tomita}}, \ and\ \bibinfo {author} {\bibfnamefont
  {T.}~\bibnamefont {Soga}},\ }\href@noop {} {\bibfield  {journal} {\bibinfo
  {journal} {Journal of chromatography A}\ }\textbf {\bibinfo {volume}
  {1159}},\ \bibinfo {pages} {134} (\bibinfo {year} {2007})}\BibitemShut
  {NoStop}%
\bibitem [{\citenamefont {Zhu}\ \emph {et~al.}(2003)\citenamefont {Zhu},
  \citenamefont {Phalakornkule}, \citenamefont {Ghosh}, \citenamefont
  {Grossmann}, \citenamefont {Koepsel}, \citenamefont {Ataai},\ and\
  \citenamefont {Domach}}]{zhu2003metabolic}%
  \BibitemOpen
  \bibfield  {author} {\bibinfo {author} {\bibfnamefont {T.}~\bibnamefont
  {Zhu}}, \bibinfo {author} {\bibfnamefont {C.}~\bibnamefont {Phalakornkule}},
  \bibinfo {author} {\bibfnamefont {S.}~\bibnamefont {Ghosh}}, \bibinfo
  {author} {\bibfnamefont {I.}~\bibnamefont {Grossmann}}, \bibinfo {author}
  {\bibfnamefont {R.}~\bibnamefont {Koepsel}}, \bibinfo {author} {\bibfnamefont
  {M.}~\bibnamefont {Ataai}}, \ and\ \bibinfo {author} {\bibfnamefont
  {M.}~\bibnamefont {Domach}},\ }\href@noop {} {\bibfield  {journal} {\bibinfo
  {journal} {Metabolic engineering}\ }\textbf {\bibinfo {volume} {5}},\
  \bibinfo {pages} {74} (\bibinfo {year} {2003})}\BibitemShut {NoStop}%
\bibitem [{\citenamefont {Zamboni}\ \emph {et~al.}(2009)\citenamefont
  {Zamboni}, \citenamefont {Fendt}, \citenamefont {R{\"u}hl},\ and\
  \citenamefont {Sauer}}]{zamboni200913c}%
  \BibitemOpen
  \bibfield  {author} {\bibinfo {author} {\bibfnamefont {N.}~\bibnamefont
  {Zamboni}}, \bibinfo {author} {\bibfnamefont {S.-M.}\ \bibnamefont {Fendt}},
  \bibinfo {author} {\bibfnamefont {M.}~\bibnamefont {R{\"u}hl}}, \ and\
  \bibinfo {author} {\bibfnamefont {U.}~\bibnamefont {Sauer}},\ }\href@noop {}
  {\bibfield  {journal} {\bibinfo  {journal} {Nature protocols}\ }\textbf
  {\bibinfo {volume} {4}},\ \bibinfo {pages} {878} (\bibinfo {year}
  {2009})}\BibitemShut {NoStop}%
\bibitem [{\citenamefont {Basan}\ \emph {et~al.}(2015)\citenamefont {Basan},
  \citenamefont {Hui}, \citenamefont {Okano}, \citenamefont {Zhang},
  \citenamefont {Shen}, \citenamefont {Williamson},\ and\ \citenamefont
  {Hwa}}]{basan2015overflow}%
  \BibitemOpen
  \bibfield  {author} {\bibinfo {author} {\bibfnamefont {M.}~\bibnamefont
  {Basan}}, \bibinfo {author} {\bibfnamefont {S.}~\bibnamefont {Hui}}, \bibinfo
  {author} {\bibfnamefont {H.}~\bibnamefont {Okano}}, \bibinfo {author}
  {\bibfnamefont {Z.}~\bibnamefont {Zhang}}, \bibinfo {author} {\bibfnamefont
  {Y.}~\bibnamefont {Shen}}, \bibinfo {author} {\bibfnamefont {J.~R.}\
  \bibnamefont {Williamson}}, \ and\ \bibinfo {author} {\bibfnamefont
  {T.}~\bibnamefont {Hwa}},\ }\href@noop {} {\bibfield  {journal} {\bibinfo
  {journal} {Nature}\ }\textbf {\bibinfo {volume} {528}},\ \bibinfo {pages}
  {99} (\bibinfo {year} {2015})}\BibitemShut {NoStop}%
\end{thebibliography}%

\end{document}